\documentclass[12pt]{article}

\usepackage{epsf,amsfonts,hyperref}
\renewcommand{\appendix}[1]{
    \setcounter{equation}{0}
    \renewcommand{\thesection}{\Alph{section}}
    \section{Appendix: \protect\indent #1}
}

\newcommand\encadremath[1]{\vbox{\hrule\hbox{\vrule\kern8pt
\vbox{\kern8pt \hbox{$\displaystyle #1$}\kern8pt}
\kern8pt\vrule}\hrule}}
\def\enca#1{\vbox{\hrule\hbox{
\vrule\kern8pt\vbox{\kern8pt \hbox{$\displaystyle #1$}
\kern8pt} \kern8pt\vrule}\hrule}}

\newcommand\figureframex[3]{
\begin{figure}[bth]
\hrule\hbox{\vrule\kern8pt
\vbox{\kern8pt \vbox{
\begin{center}
{\mbox{\epsfxsize=#1.truecm\epsfbox{#2}}}
\end{center}
\caption{#3}
}\kern8pt}
\kern8pt\vrule}\hrule
\end{figure}
}
\newcommand\figureframey[3]{
\begin{figure}[bth]
\hrule\hbox{\vrule\kern8pt
\vbox{\kern8pt \vbox{
\begin{center}
{\mbox{\epsfysize=#1.truecm\epsfbox{#2}}}
\end{center}
\caption{#3}
}\kern8pt}
\kern8pt\vrule}\hrule
\end{figure}
}

\makeatletter
\@addtoreset{equation}{section}
\makeatother
\newtheorem{theorem}{Theorem}[section]

\newtheorem{remark}{Remark}[section]
\newtheorem{proposition}{Proposition}[section]
\newtheorem{lemma}{Lemma}[section]
\newtheorem{corollary}{Corollary}[section]
\newtheorem{definition}{Definition}[section]
\def\br{\begin{remark}\rm\small}
\def\er{\end{remark}}
\def\bt{\begin{theorem}}
\def\et{\end{theorem}}
\def\bd{\begin{definition}}
\def\ed{\end{definition}}
\def\bp{\begin{proposition}}
\def\ep{\end{proposition}}
\def\bl{\begin{lemma}}
\def\el{\end{lemma}}
\def\bc{\begin{corollary}}
\def\ec{\end{corollary}}
\def\beaq{\begin{eqnarray}}
\def\eeaq{\end{eqnarray}}
\newcommand{\proof}[1]{{\noindent \bf proof:}\par
{#1} $\square$}

\newcommand{\eq}[1]{eq.~(\ref{#1})}

\newcommand{\beq}{\begin{equation}}
\newcommand{\eeq}{\end{equation}}
\newcommand{\bea}{\begin{eqnarray}}
\newcommand{\eea}{\end{eqnarray}}

\renewcommand{\and}{{\qquad {\rm and} \qquad}}

\newcommand{\virg}{{\qquad , \qquad}}

 \newcommand{\Tr}{{\,\rm Tr}\:}

\newcommand{\td}[1]{{\tilde{#1}}}
\newcommand{\ovl}[1]{{\overline{#1}}}

\renewcommand{\l}{\lambda}
\newcommand{\om}{\omega}

\newcommand{\ee}[1]{{{\rm e}^{#1}}}

\newcommand{\Pint}{{\int\kern -1.em -\kern-.25em}}

\newcommand{\bcycle}{{\cal B}}
\newcommand{\acycle}{{\cal A}}
\newcommand{\Ycl}{{Y_{\rm cl}}}
\newcommand{\npole}{{\overline{n} }}
\newcommand{\genus}{{\overline{g}}}

\renewcommand\l{\lambda}

\newcommand\Res{\mathop{{\rm Res}}}

\begin{document}
\sloppy


\pagestyle{empty}
\hfill SPhT-T08/140
\addtolength{\baselineskip}{0.20\baselineskip}
\begin{center}
\vspace{26pt}
{\large \bf {Topological expansion of the Bethe ansatz,\\ and non-commutative algebraic geometry}}
\newline
\vspace{26pt}

{\sl B.\ Eynard}${}^\dagger$\hspace*{0.05cm}\footnote{ E-mail: bertrand.eynard@cea.fr },
{\sl O.\ Marchal}${}^\dagger\, {}^\ddagger$\hspace*{0.05cm}\footnote{ E-mail: olivier.marchal@cea.fr }\\
\vspace{6pt}
${}^\dagger$ 
Institut de Physique Th\'eorique,\\
CEA, IPhT, F-91191 Gif-sur-Yvette, France,\\
CNRS, URA 2306, F-91191 Gif-sur-Yvette, France.\\
${}^\ddagger$ Centre de recherches math\'ematiques, Universit\'e de Montr\'eal 
C. P. 6128, succ. centre ville, Montr\'eal, Qu\'ebec, Canada H3C 3J7.
\end{center}

\vspace{20pt}
\begin{center}
{\bf Abstract}:

In this article, we define a non-commutative deformation of the "symplectic invariants" (introduced in \cite{EOFg}) of an algebraic hyperelliptical plane curve. The necessary condition for our definition to make sense is a Bethe ansatz. The commutative limit reduces to  the symplectic invariants, i.e. algebraic geometry, and thus we define non-commutative deformations of some algebraic geometry quantities. 
In particular our non-commutative Bergmann kernel satisfies a Rauch variational formula.
Those non-commutative invariants are inspired from the large N expansion of formal non-hermitian matrix models. Thus they are expected to be related to the enumeration problem of discrete non-orientable surfaces of arbitrary topologies.

\end{center}


\vspace{26pt}
\pagestyle{plain}
\setcounter{page}{1}


\section{Introduction}

In \cite{EOFg}, the notion of symplectic invariants of a spectral curve was introduced. For any given  algebraic plane curve (called spectral curve) of equation:
\beq
0={\cal E}(x,y)= \sum_{i,j} {\cal E}_{i,j}\,\, x^i\, y^j
\eeq
an infinite sequence of numbers
\beq
F^{(g)}({\cal E})
\qquad , \,\, g=0,1,2,\dots,\infty
\eeq
and an infinite sequence of multilinear meromorphic forms $W_n^{(g)}$ (meromorphic on the algebraic Riemann surface of equation ${\cal E}(x,y)=0$) were defined.

Their definition was inspired from hermitian matrix models, i.e. in the case where ${\cal E}={\cal E}_{\rm M.M.}$ is the spectral curve ($y(x)$ is the equilibrium density of eigenvalues) of a formal hermitian matrix integral $Z_{\rm M.M.}=\int dM\, \ee{-N\Tr V(M)}$, the $F^{(g)}$ were such that:
\beq
\ln{Z_{\rm M.M.}} = \sum_{g=0}^\infty N^{2-2g} F^{(g)}({\cal E}_{\rm M.M.})
\eeq
The $F^{(g)}$'s have many remarkable properties (see \cite{EOFg}), in particular invariance under symplectic deformations of the spectral curve, homogeneity (of degree $2-2g$), holomorphic anomaly equations (modular transformations), stability under singular limits, ...
An important property also, is that the following formal series
\beq
\tau({\cal E})= \ee{\sum_g N^{2-2g} F^{(g)}({\cal E})}
\eeq
is the "formal" $\tau$ function of an integrable hierarchy.

\smallskip
Although those notions were first developed for matrix models, they extend beyond matrix models, and they make sense for spectral curves which are not matrix models spectral curves.
For instance the (non-algebraic) spectral curve ${\cal E}_{\rm WP}(x,y) = (2\pi y)^2 - (\sin{(2\pi \sqrt{x})})^2$ is such that $F^{(g)}({\cal E}_{\rm WP})={\rm Vol}(\overline{\cal M}_{g})$ is the Weyl-Petersson volume of moduli space of Riemann surfaces of genus $g$ (see \cite{EynVolmum, EOVolWP}).
It is conjectured \cite{remodelBmodel} that the $F^{(g)}$'s are deeply related to Gromov-Witten invariants, Hurwitz numbers \cite{BMHurw} and topological strings \cite{remodelBmodel}. In particular they are related to the Kodaira-Spencer field theory \cite{DVKS}.

\bigskip

There were many attempts to compute also non-hermitian matrix integrals, and an attempt to extend the method of \cite{EOFg} was first made in \cite{ChekEynbeta}, and here in this paper we deeply improve the result of \cite{ChekEynbeta}.
The aim of the construction we present here, is to define $F^{(g)}$'s for a "non-commutative spectral curve", i.e. a non commutative polynomial:
\beq
{\cal E}(x,y) = \sum_{i,j} {\cal E}_{i,j}\,\, x^i\, y^j
\virg
[y,x]=\hbar
\eeq
For instance we can view $y$ as $y=\hbar\, {\partial/\partial x}$, and ${\cal E}$ is a differential operator, which encodes a linear differential equation.

In this article we choose ${\cal E}(x,y)$ of degree 2 in the variable $y$, i.e. the case of a second order linear differential equation, i.e. Schroedinger equation, and we leave to a further work the general case.

\bigskip
Here, in this article, we define some $F^{(g)}({\cal E})$, which reduce to those of \cite{EOFg} in the limit $\hbar\to 0$, and which compute non-hermitian matrix model topological expansions.

For instance consider a formal matrix integral:
\beq
Z=\int_{E_{2\beta,N}} dM \ee{-N\sqrt\beta \Tr V(M)} = \ee{\sum_g N^{2-2g}\, F^{(g)}}
\eeq
where $E_{2\beta,N}$ is one of the Wigner matrix ensembles \cite{Mehtabook} of rank $N$: $E_{1,N}$ is the set of real symmetric matrices, $E_{2,N}$ is the set of hermitian matrices, and $E_{4,N}$ is the set of self-dual quaternion matrices (see \cite{Mehtabook} for a review).
We define:
\beq
\hbar={1\over N}\left(\sqrt\beta - {1\over \sqrt\beta}\right)
\eeq
Notice that $\hbar=0$ for hermitian matrices, i.e. the hermitian case is the classical limit $[y,x]=0$.
Notice also that the expected duality $\beta\leftrightarrow 1/\beta$ (cf \cite{mkrtchyan1, Bryc}) corresponds to $\hbar\leftrightarrow -\hbar$, i.e. we expect it to correspond to the duality $x\leftrightarrow y$ (for $\hbar=0$, the $x\leftrightarrow y$ duality was proved in \cite{EOsymxy}).

Let us also mention that the topological expansion of non-hermitian matrix integrals is known to be related to the enumeration of unoriented discrete surfaces, and we expect that our $F^{(g)}=\sum_k \hbar^k \,F^{(g,k)}$ can be interpreted as generating functions of such unoriented surfaces.

So, in this article, we provide a method for computing $F^{(g,k)}$ for any $g$ and $k$ (which is more consise than  \cite{ChekEynbeta}).

\subsubsection*{Outline of the article}

\begin{itemize}
\item In section \ref{secdef}, we introduce our recursion kernel $K(x,x')$, and we show that the mere existence of this kernel is equivalent to the Bethe ansatz condition.

\item In section \ref{secdefWngFg}, we define the $W_n^{(g)}$'s and the $F^{(g)}$'s, and we study their main properties, for instance that $W_n^{(g)}$ is symmetric.

\item In section \ref{seclimitclassical}, we study the classical limit $\hbar\to 0$, and we show that we recover the algebro-geometric construction of \cite{EOFg}.

\item This inspires a notion of non-commutative algebraic geometry in section \ref{secqalgeo}.

\item In section \ref{secMM}, we study the application to the topological expansion of non-hermitian matrix integrals.

\item In section \ref{secGaudin}, we study the application to the Gaudin model.


\item Section \ref{secConcl} is the conclusion.

\item All the technical proofs are written in appendices for readability.

\end{itemize}

\section{Definitions, kernel and Bethe ansatz}\label{secdef}

Let $V'(x)$ be a rational function (possibly a polynomial), and we call $V(x)$ the {\bf potential}.
Let $\alpha_i$ be the poles of $V'(x)$ (one of the poles may be at $\infty$). 

\medskip For example, the following potential is called {\bf Gaudin potential} (see section \ref{secGaudin}):
\beq
V_{\rm Gaudin}'(x) = x+ \sum_{i=1}^{\npole} {S_i\over x-\alpha_i}
\eeq
As another example, we will consider formal matrix models in section \ref{secMM}, for which
$V'(x)$ is a polynomial.

However, many other choices can be made.

\subsection{The problem}

Our problem is to find $m$ complex numbers $s_1,\dots, s_m$, as well as two functions $G(x_0,x)$ and $K(x_0,x)$ with the following properties:
\begin{enumerate}
\item $G(x_0,x)$ is a rational function of $x$ with poles at $x=s_i$, and a simple pole of residue $+1$ at $x=x_0$, and which behaves as $O(1/x)$ at $x\to\infty$.
\item $G(x_0,x)$ is a rational function of $x_0$ with (possibly multiple) poles at $x_0=s_i$, and a simple pole at $x_0=x$, and $G(x_0,x)$ behaves like $O(1/x_0)$ at $x_0\to\infty$.
\item $B(x_0,x)=-{1\over 2} {\partial\over \partial x}G(x_0,x)$ is symmetric:
$B(x_0,x)=B(x,x_0)$.
\item $K$ and $G$ are related by the following differential equation:
\beq\label{diffeqdefK}
\left(2\hbar \sum_{i=1}^m {1\over x-s_i} - V'(x) - \hbar {\partial\over \partial x}\right)\, K(x_0,x) = G(x_0,x)
\eeq
\item $K(x_0,x)$ is analytical when $x\to s_i$ for all $i=1,\dots,m$.

\end{enumerate}

We shall see below that those 5 conditions determine $K$, $G$, and the $s_i$'s.
In fact condition 5 is the most important one in this list, it amounts to a {\bf no-monodromy condition}, and we shall see below that it implies that the $s_i$'s must obey the {\bf Bethe-ansatz equation}.

\subsection{Analytical structure of the kernel $G$}

The 4th and 5th conditions imply that $G(x_0,x)$ has at most simple poles at $x=s_i$. Then condition 3 implies that $G(x_0,x)$ has at most double poles at $x_0=s_i$.

The first 3 conditions imply that there exists a symmetric matrix $A_{i,j}$ such that $G(x_0,x)$ can be written:
\beq
G(x_0,x) = {1\over x-x_0} + 2\sum_{i,j=1}^m {A_{i,j}\over (x-s_i)(x_0-s_j)^2} 
\eeq
and therefore:
\beq
B(x_0,x) = {1\over 2}\,{1\over (x-x_0)^2} + \sum_{i,j=1}^m {A_{i,j}\over (x-s_i)^2(x_0-s_j)^2} 
\eeq
We will argue in section \ref{secqalgeo}, that $B$ can be viewed as a non=commutative deformation of the algebraic geometry's Bergmann kernel.

\subsection{Bethe ansatz and monodromies}\label{sectbetheansatz}

First, we study the conditions under which the differential equation \eq{diffeqdefK} has no monodromies around $s_i$, in other words the condition under which $K(x_0,x)$ is analytical when $x\to s_i$, $\forall i$:
\beq
K(x_0,s_i+\epsilon) = K(x_0,s_i)+\epsilon K'(x_0,s_i)+{\epsilon^2\over 2} K''(x_0,s_i)+{\epsilon^3\over 6} K'''(x_0,s_i)+\dots
\eeq

\medskip
Equating the coefficient of $\epsilon^{-1}$ in \eq{diffeqdefK}, we get:
\beq\label{eqmono1}
\hbar K(x_0,s_i) =  \sum_{j} {A_{i,j}\over (x_0-s_j)^2}
\eeq
equating the coefficient of $\epsilon^{0}$ in \eq{diffeqdefK}, we get:
\beq\label{eqmono2}
\hbar K'(x_0,s_i) = {-1\over x_0-s_i} + V'(s_i) K(x_0,s_i) - 2\hbar\sum_{j\neq i} {K(x_0,s_i)-K(x_0,s_j)\over s_i-s_j} 
 \eeq
and equating the coefficient of $\epsilon^{1}$ in \eq{diffeqdefK}, we get:
\bea\label{eqmono3}
&& 2\hbar \sum_{j\neq i}{K'(x_0,s_i)\over s_i-s_j} -2\hbar \sum_{j\neq i}{K(x_0,s_i)\over (s_i-s_j)^2} +V''(s_i)K(x_0,s_i)\cr
&=& V'(s_i) K'(x_0,s_i) - {1\over (s_i-x_0)^2} - 2 \sum_{j\neq i} \sum_{k} {A_{j,k}\over (s_i-s_j)^2(x_0-s_k)^2} \cr
\eea
Notice from \eq{eqmono1}, that $K(x_0,s_i)$ has only double poles in $x_0$, with no residue:
\beq
\Res_{x_0\to s_k} K(x_0,s_i)=0
\eeq
Then, taking the residue at $x_0\to s_k$ in \eq{eqmono2}, we see that:
\beq
\hbar \Res_{x_0\to s_k} K'(x_0,s_i) = -\delta_{i,k} 
\eeq
Then, 
taking the residue when $x_0\to s_i$ in \eq{eqmono3}, implies that the $s_i$'s are Bethe roots, i.e. they must obey the {\bf Bethe equation}:
\beq\label{Betheeq}
\encadremath{
\forall\, i=1,\dots,m\, , \qquad \quad 2\hbar\,\sum_{j\neq i} {1\over s_i-s_j} = V'(s_i)
}\eeq
Then \eq{eqmono3} becomes:
\beq\label{defA}
{1\over (s_i-x_0)^2}
=  V''(s_i)K(x_0,s_i) + 2\hbar \sum_{j\neq i}{K(x_0,s_i)\over (s_i-s_j)^2}    - 2 \sum_{j\neq i} \sum_{k} {A_{j,k}\over (s_i-s_j)^2(x_0-s_k)^2}
\eeq
i.e. by comparing the coefficient of $1/(x_0-s_k)^2$ on both sides:
\beq\label{Anomono}
\encadremath{
\delta_{i,k} = {1\over \hbar}V''(s_i) A_{i,k} + 2 \sum_{j\neq i} {A_{i,k}-A_{j,k}\over (s_i-s_j)^2}
}\eeq
i.e. $A$ is the inverse of the Hessian matrix $T$:
\beq
A=T^{-1}
\virg
\left\{\begin{array}{l}
T_{i,i} = {1\over \hbar}V''(s_i) + 2 \sum_{j\neq i} {1\over (s_i-s_j)^2} \cr
T_{i,j} = - \, {2\over (s_i-s_j)^2} \cr
\end{array}\right.
\eeq
\beq
T_{i,j} = {1\over \hbar}\,\,{\partial^2\over \partial s_i \partial s_j}\,\,\Big(  \sum_k V(s_k) - \hbar \sum_{k\neq l} \ln{(s_k-s_l)}\Big)
\eeq

Therefore the Bethe ansatz equations \eq{Betheeq} (as well as \eq{Anomono}) are the necessary conditions for $K(x_0,x)$ to be analytical when $x\to s_i$. Those conditions are necessary, but also sufficient conditions, as one can see by solving explicitely the linear ODE for $K$.

\beq\label{solODEK}
K(x_0,x) =  \int^x_{c}\,dx' G(x_0,x')\,\,\ee{{1\over \hbar} (V(x')-V(x))}\,\,\prod_{i} {(x-s_i)^2\over (x'-s_i)^2}
\eeq

\medskip

\br
Notice that $K(x_0,x)$ is not analytical everywhere, it has a logarithmic singularity  at $x=x_0$, and it has essential singularities at the poles of $V'$.
\er

\br
Notice that if one solution of the ODE is analytical near all $s_i$'s, then all solutions have that property.
Indeed, all the solutions differ by a solution of the homogeneous equation, i.e. by:
\beq
\prod_{i} (x-s_i)^2\,\, \ee{-{1\over \hbar} V(x)}
\eeq
which is clearly analytical near the $s_i$'s.

So, for the moment, the requirements 1--5 determine $G(x_0,x)$ uniquely, but $K(x_0,x)$ is not unique.
Let us choose one possible $K(x_0,x)$, and we prove below in theorem \ref{thWngindeptK}, that the objects we are going to define, do not depend on the choice of $K$.

\er

\br
In what follows, it is useful to compute the Taylor expansion of $K$ near a root $s_i$. We write:
\beq
K(x_0,x) = \sum_{k=0}^\infty K_{i,k}(x_0)\,\,(x-s_i)^k
\eeq
The coefficients $K_{i,k}(x_0)$ are themselves rational fractions of $x_0$, and are computed in appendix \ref{appKexpansion}.
\er

\subsection{Schroedinger equation}

It is well known  that the Bethe condition can be rewritten as a Schroedinger equation \cite{BBTbook, BabBetGaudin}. We rederive it here for completeness.

Define the wave function:
\beq
\psi(x) = \prod_{i=1}^m (x-s_i) \,\,\, \ee{-{1\over 2\hbar}\, V(x)}
\virg
\om(x) = \hbar \sum_{i=1}^m {1\over x-s_i}
\eeq
\beq
Y(x) = -2 \hbar {\psi'(x)\over \psi(x)} = V'(x)-2\om(x) = V'(x) - 2\hbar \sum_i {1\over x-s_i}
\eeq
then compute:
\bea\label{eqRicattiUY}
U(x) 
&=& Y^2 -2 \hbar Y'(x) = 4\hbar^2 {\psi''(x)\over \psi(x)} \cr
&=& V'(x)^2 - 2\hbar V''(x) + 4(\om(x)^2-V'(x)\om(x)+\hbar \om'(x))
\eea

We have:
\bea
\om(x)^2+\hbar \om'(x) 
&=& \hbar^2 \sum_{i,j} {1\over (x-s_i)(x-s_j)} - \hbar^2 \sum_{i} {1\over (x-s_i)^2}  \cr
&=& \hbar^2 \sum_{i\neq j} {1\over (x-s_i)(x-s_j)}  \cr
\eea
which is a rational fraction with only simple poles at the $s_i$'s.
The residue at $s_i$ is $2\hbar^2 \sum_{j\neq i} {1\over s_i-s_j} = \hbar V'(s_i)$, and thus:
\beq
\om(x)^2+\hbar \om'(x) = \hbar \sum_{i} {V'(s_i)\over (x-s_i)}  
\eeq
which implies:
\beq
\om(x)^2 - V'(x)\om(x) +\hbar \om'(x) = - \hbar \sum_{i} {V'(x)-V'(s_i)\over (x-s_i)}  
\eeq
and thus:
\beq\label{eqUV}
U(x) = V'(x)^2 - 2\hbar V''(x)  - 4 \hbar \sum_{i=1}^m {V'(x)-V'(s_i)\over x-s_i}
\eeq

Therefore $U(x)$ is a rational fraction with poles at the poles of $V'$ (of degree at most those of $V'^2$),  in particular it has no poles at the $s_i$'s.

$U$ is the potential for the Schroedinger equation for $\psi$:
\beq
\encadremath{
4\hbar^2 \psi'' =  U\, \psi
}\eeq

\medskip
As announced in the introduction, this equation can be encoded in a D-module element:
\beq
{\cal E}(x,y) = y^2-{1\over 4}U(x)
\virg
y=\hbar\,{\partial \over \partial x}
\virg [y,x]=\hbar
\eeq
i.e.
\beq
{\cal E}(x,y).\psi = 0
\eeq

Notice that the Schroedinger equation is equivalent to a Ricatti equation for $Y=-2\hbar \psi'/\psi$:
\beq
\encadremath{
Y^2 - 2\hbar Y' = U
}\eeq

\subsection{Classical limit}
\label{secclaslim1}

We shall come back in more detail to the classical limit $\hbar\to 0$ in section \ref{seclimitclassical}.
However, let us already make a few comments.

\bigskip
$\bullet$ In the classical limit, the Ricatti equation becomes an algebraic equation (hyperelliptical), which we call the (classical) spectral curve:
\beq
Y_{\rm cl}^2 = U(x) 
\eeq
The function $Y_{\rm cl}(x)=\sqrt{U(x)}$ is therefore a multivalued function of $x$, and it should be seen as a meromorphic function on a branched Riemann surface (branching points are the zeroes of $U(x)$).
We shall see below that in the limit $\hbar\to 0$, the kernel $B(x_0,x)$ tends towards the Bergmann kernel of that Riemann surface.

In other words the classical limit is expressed in terms of {\bf algebraic geometry}.

In fact, in this article we are going to define non-commutative deformations of certain algebraic geometric objects in section \ref{secqalgeo}.

\section{Definition of correlators and free energies}
\label{secdefWngFg}

In this section, we define the quantum deformations of the symplectic invariants introduced in \cite{Eyn1loop, EOFg}.
The following definitions are inspired from (not hermitian) matrix models. The special case of their application to matrix models will be discussed in section \ref{secMM}.

\subsection{Definition of correlators}

\bd\label{defWng}
We define the following functions $W_n^{(g)}(x_1,\dots,x_n)$ (called $n$-point correlation function of "genus"\footnote{here $g$ is any given integer, it has nothing to do with the genus of the spectral curve.} $g$) by the recursion:
\beq
W_1^{(0)}(x) = \om(x) = \hbar \sum_{i=1}^m {1\over x-s_i}
\virg
W_2^{(0)}(x_1,x_2)=B(x_1,x_2)
\eeq
\bea\label{mainrecformula}
&& W^{(g)}_{n+1}(x_0,J)  \cr
&=&   \sum_{i=1}^m \Res_{x\to s_i}  K(x_0,x)\, \left( \ovl{W}_{n+2}^{(g-1)}(x,x,J) + \sum_{h=0}^g \sum'_{I\subset J} W_{|I|+1}^{(h)}(x,I) W_{n-|I|+1}^{(g-h)}(x,J/I) \right) \cr
\eea
where $J$ is a collective notation for the variables $J=\{ x_{1},\dots,x_{n} \}$, and where $\sum\sum'$ means that we exclude the terms $(h=0,I=\emptyset)$ and $(h=g,I=J)$, and where:
\beq
\ovl{W}_{n}^{(g)}(x_1,...,x_n) = W_{n}^{(g)}(x_1,...,x_n) - {\delta_{n,2}\delta_{g,0}\over 2}\, {1\over (x_1-x_2)^2}
\eeq

\ed

\vspace{0.5cm}
\br
This is exactly the same recursion as in \cite{EOFg}, the only difference is that the kernel $K$ is not algebraic, but it is solution of the differential equation \eq{diffeqdefK}.
We shall show in section \ref{seclimitclassical}, that in the limit $\hbar\to 0$, it indeed reduces to the definition of \cite{EOFg}.
\er

\br
We say that $W_n^{(g)}$ is the correlation function of genus $g$ with $n$ marked points, and sometimes we say that it has characteristics:
\beq
\chi=2-2g-n
\eeq
By analogy with algebraic geometry, we say that $W_n^{(g)}$ is stable if $\chi<0$ and unstable if $\chi\geq 0$.
We see that all the stable $W_n^{(g)}$'s have a common recursive definition  def.\ref{defWng}, whereas the unstable ones appear as exceptions.
\er

\br
In order for the definition to make sense, we must make sure that the behaviour of each term in the vicinity of $x\to s_i$ is indeed locally meromorphic so that we can compute residues, i.e. there must be no log-singularity near $s_i$.
In particular, the requirement of section \ref{sectbetheansatz} for the kernel $K$ is {\bf necessary}.
In other words, a necessary condition for definition eq.\ref{mainrecformula} to make sense, is the {\bf Bethe ansatz} !
\er

\subsection{Properties of correlators}

The main reason of definition. \ref{defWng}, is because the $W_n^{(g)}$'s have many beautiful properties, which generalize those of \cite{EOFg}.

We shall prove the following properties:

\bt\label{thpolessiWng}
Each $W_n^{(g)}$ is a rational function of all its arguments. It has poles only at the $s_i$'s (except $W_2^{(0)}$, which also has a pole at $x_1=x_2$).
In particular it has no poles at the $\alpha_i$'s.
Moreover, it vanishes as $O(1/x_i)$ when $x_i\to\infty$.
\et
\proof{ in appendix \ref{approofthpolessiWng}}

\bt\label{thWngPng}
 The $W_n^{(g)}$'s satisfy the loop equation, i.e. Virasoro-like constraints.
This means that the quantity:
\bea\label{loopeqPng}
 P_{n+1}^{(g)}(x;x_1...,x_n)
 &=&
-Y(x)\overline{W}_{n+1}^{(g)}(x,x_1,...,x_n) + \hbar \partial_{x}{\overline{W}_{n+1}^{(g)}(x,x_1...,x_n)} \cr
&& + \sum_{I\subset J} \ovl{W}_{|I|+1}^{(h)}(x,x_I) \ovl{W}_{n-|I|+1}^{(g-h)}(x,J/I) +
\ovl{W}_{n+2}^{(g-1)}(x,x,J)  \cr
& &+ \sum_{j}
\partial_{x_j} \left( {{\ovl{W}_n^{(g)}(x,J/\{j\})-{\ovl{W}_n^{(g)}(x_j,J/\{j\})}} \over {(x-x_j)}}\right) \cr
\eea
is a rational fraction of $x$ (possibly a polynomial), with no pole at $x=s_i$.
The only possible poles of $P_{n+1}^{(g)}(x;x_1...,x_n)$ are at the poles of $V'(x)$, with degree less than the degree of $V'$.

\et
\proof{ in appendix \ref{approofthWngPng}}

\bt\label{thsym}
 Each $W_n^{(g)}$ is a symmetric function of all its arguments.
\et
\proof{ in appendix \ref{approofthsym}, with the special case of $W_3^{(0)}$ in appendix \ref{approofthW3Krich}.}

\bt\label{thWngindeptK}
The correlation functions $W_n^{(g)}$ are independent of the choice of kernel $K$, provided that $K$ is solution of the equation \eq{diffeqdefK}.
\et
\proof{ in appendix \ref{approofthWngindeptK}}


\bt\label{thW3Krich}
The 3 point function $W_3^{(0)}$ can also be written:
\beq
W_3^{(0)}(x_1,x_2,x_3) = 4\,\sum_i \Res_{x\to s_i}\,\, {B(x,x_1)B(x,x_2)B(x,x_3)\over Y'(x)}
\eeq
(In section \ref{secqalgeo}, we interpret this equation as a non-commutative version of Rauch variational formula).
\et
\proof{ in appendix \ref{approofthW3Krich}}

\bt\label{thvariationV}
Under an infinitesimal variation of the potential $V\to V+\delta V$, we have:
\beq
\forall n\geq 0, g\geq 0
\,\, , \quad
\delta W_{n}^{(g)}(x_1,\dots,x_n) = - \sum_i \Res_{x\to s_i} W_{n+1}^{(g)}(x,x_1,\dots,x_n)\, \delta V(x)
\eeq
\et
\proof{ in appendix \ref{approofthvariationV}}

This theorem suggest the definition of the "loop operator":
\bd
The loop operator $\delta_x$ computes the variation of $W_n^{(g)}$ under a formal variation $\delta_x V(x')={1\over x-x'}$:
\beq
\delta_{x_{n+1}}\, W_n^{(g)}(x_1,\dots,x_n) = W_{n+1}^{(g)}(x_1,\dots,x_n,x_{n+1})
\eeq
The loop operator is a derivation: $\delta_x (uv) = u\delta_x v+v\delta_x u$, and we have $\delta_{x_1}\delta_{x_2} = \delta_{x_2}\delta_{x_1}$, $\delta_{x_1} \partial_{x_2} =\partial_{x_2}  \delta_{x_1}$.
\ed

\bt\label{thResY} 
For $n\geq 1$, $W_n^{(g)}$ satify the equation:
\beq
 \sum_{i=1}^n {\partial \over \partial x_i}\, \ovl{W}_n^{(g)}(x_1,\dots,x_n)  
= -\sum_i \Res_{x_{n+1}\to s_i}\,\, V'(x_{n+1})\,\,\ovl{W}_{n+1}^{(g)}(x_1,\dots,x_n,x_{n+1})
\eeq
and
\beq
 \sum_{i=1}^n {\partial \over \partial x_i}\, x_i \,\ovl{W}_n^{(g)}(x_1,\dots,x_n)  
= -\sum_i \Res_{x_{n+1}\to s_i}\,\, x_{n+1}\,V'(x_{n+1})\,\,\ovl{W}_{n+1}^{(g)}(x_1,\dots,x_n,x_{n+1})
\eeq
\et
\proof{ in appendix \ref{approofthResY}}

\bt\label{thdilaton} 
For $n\geq 1$, $W_n^{(g)}$ satify the equation:
\beq
(2-2g-n-\hbar {\partial\over \partial \hbar})\, \ovl{W}_{n}^{(g)}(x_1,\dots,x_n)
= - \sum_i \Res_{x_{n+1}\to s_i}\,\, V(x_{n+1})\,\,\ovl{W}_{n+1}^{(g)}(x_1,\dots,x_n,x_{n+1})
\eeq
\et
\proof{ We give a "long" proof in appendix \ref{approofthdilaton}.

There is also a short cut:

If one changes $\hbar\to \l \hbar$, and $V\to \l V$, the $s_i$'s don't change, $B$ and $G$ don't change, and $K$ changes to ${1\over \l}\,K$, thus $W_n^{(g)}$ changes by $\l^{2-2g-n} W_n^{(g)}$.
The theorem is obtained by computing ${\l\partial\over \partial \l} \l^{2g-2+n} W_n^{(g)} = \sum_k {t_k\partial\over \partial t_k} W_n^{(g)} $, and computing the RHS with theorem \ref{thvariationV}, i.e. $\delta V=V$. 

}

\subsection{Definition of free energies}

So far, we have defined $W_n^{(g)}$ with $n\geq 1$.
Now, we define $F^{(g)}=W_0^{(g)}$.
\bigskip

Theorem \ref{thvariationV}, and the symmetry theorem \ref{thsym} imply that:
\beq
\delta_{x_1} W_1^{(g)}(x_2) = W_2^{(g)}(x_1,x_2) = W_2^{(g)}(x_2,x_1)
= \delta_{x_2} W_1^{(g)}(x_1) 
\eeq
Thus, the symmetry of $W_2^{(g)}$ implies that there exists a "free energy" $F^{(g)}=W_0^{(g)}$ such that:
\beq
W_{1}^{(g)}(x) = \delta_x F^{(g)} 
\eeq
which is equivalent to saying that for any variation $\delta V$:
\beq
\delta F^{(g)} = - \sum_i \Res_{x\to s_i} W_{1}^{(g)}(x)\, \delta V(x)
\eeq
Therefore, we know that there must exists some $F^{(g)}=W_0^{(g)}$ which satisfy theorem \ref{thvariationV} for $n=0$.

\bigskip

Now, let us give a definition of $F^{(g)}$, inspired from theorem \ref{thdilaton}, and which will be proved to satisfy theorem \ref{thvariationV} for $n=0$.

\bd\label{defallFg}
We define $F^{(g)} \equiv W_0^{(g)}$ by a solution of the differential equation in $\hbar$:
\beq
\forall g\geq 2 \virg
(2-2g-\hbar {\partial\over \partial \hbar})\, F^{(g)} =  - \sum_i \Res_{x\to s_i}\,\, W_{1}^{(g)}(x) \, V(x)
\eeq
more precisely:
\beq\label{defFg}
F^{(g)} =  \hbar^{2-2g}\,\int_0^{\hbar}\, {d\td{\hbar}\over {\td\hbar}^{3-2g}}\,\, \sum_i \Res_{x\to s_i}\,\, V(x)\,\,\, \left. W_{1}^{(g)}(x)\right|_{{\td\hbar}}
\eeq

And the unstable cases $2-2g\geq 0$ are defined by:
\beq
F^{(0)} =   \hbar^2 \sum_{i\neq j} \ln{(s_i-s_j)}  - \hbar \sum_i V(s_i)
\eeq
\beq
F^{(1)} = {1\over 2} \ln{\det A} + \ln{(\Delta(s)^2)} + {F^{(0)}\over \hbar^2} 
\eeq
where $\Delta(s) = \prod_{i>j} (s_i-s_j)$ is the Vandermonde determinant of the $s_i$'s.
\ed

\bigskip

{\bf Properties of the $F^{(g)}$'s:}

The definition of the $F^{(g)}$'s, is made so that all the theorems for the $W_n^{(g)}$'s, hold for for $n=0$ as well.
Proofs are given in appendices \ref{approofFg}, \ref{approofF0}, \ref{approofF1}.

\bigskip

Explicit computations of the first few $F^{(g)}$'s are given in section \ref{secGaudin} and appendix \ref{appm1}.

\section{Classical limit and WKB expansion}
\label{seclimitclassical}

In the $\hbar\to 0$ limit, all quantities can be expanded formally into powers of $\hbar$:
Write:
\beq
W_n^{(g)}(x_1,\dots,x_n) = \sum_k \hbar^k W_n^{(g,k)}(x_1,\dots,x_n)
\virg
F^{(g)} = \sum_k \hbar^k F^{(g,k)}
\eeq

\subsection{Classical limit}

Here we consider the classical limit $\hbar\to 0$.
We noticed in section \ref{secclaslim1}, that in that limit, the Ricatti equation
\beq
Y^2 -2\hbar Y' = U= V'^2 -2\hbar V'' - 4P
\eeq
where $P(x) = \hbar\sum_i {V'(x)-V'(s_i)\over x-s_i}$,
becomes an algebraic hyperelliptical equation:
\beq
\Ycl^2 = U(x) = V'(x)^2 - 4P(x)
\eeq
i.e.
\beq
Y(x) \mathop{\sim}_{\hbar\to 0} \Ycl(x) =  \sqrt{V'(x)^2-4P(x)}
\eeq
$\Ycl(x)$ is a multivalued function of $x$, and it should be seen as a meromorphic function on a 2-sheeted Riemann surface, i.e. there is a Riemann surface $\Sigma$ (of equation $0={\cal E}_{\rm cl}(x,y)=y^2-4U(x)$, such that the solutions of ${\cal E}_{\rm cl}(x,y)=0$ are parametrized by two meromorphic functions on $\Sigma$:
\beq
{\cal E}_{\rm cl}(x,y)=0
\quad\Leftrightarrow\quad
\exists z\in \Sigma \left\{
\begin{array}{l}
x=x(z) \cr
y=y(z)
\end{array}\right.
\eeq

The Riemann surface $\Sigma$ has a certain topology\footnote{This genus $\genus$ has nothing to do with the index $g$ of $F^{(g)}$ or $W_n^{(g)}$. } characterized by its genus $\genus$.
It has a (non-unique) symplectic basis of $2\genus$ non-trivial cycles $\acycle_i\cap \bcycle_j=\delta_{i,j}$.

The meromorphic forms on $\Sigma$ are classified as 1st kind (no pole), 3rd kind (only simple poles), and 2nd kind (multiple poles without residues).

There exists a unique 2nd kind differential $B_{\rm cl}$ on $\Sigma$, called the Bergmann kernel, such that:
$B_{\rm cl}(z_1,z_2)$ has a double pole at $z_1\to z_2$, and no other pole, without residue and normalized (in any local coordinate $z$) as:
\beq
B_{\rm cl}(z_1,z_2) \mathop{\sim}_{z_2\to z_1} {dz_1 dz_2\over (z_1-z_2)^2}+ {\rm reg}
\virg
\forall i=1,\dots,\genus\, , \,\,\, \oint_{\acycle_i} B_{\rm cl} = 0
\eeq
We define a primitive:
\beq
G_{\rm cl}(z_0,z) = -2 \int^z B_{\rm cl}(z_0,z')
\eeq
which is a 3rd kind differential in the variable $z_0$, it is called $dE_{z}(z_0)$ in \cite{EOFg}.

When $\hbar=0$, the kernel $K(z_0,z)$ satisfies the equation:
\beq
K_{\rm cl}(z_0,z) = -\, {G_{\rm cl}(z_0,z)\over Y_{\rm cl}(z)} = 2\, {\int_c^z B_{\rm cl}(z_0,z')\over Y_{\rm cl}(z)}
\eeq
which coincides with the definition of the recursion kernel in \cite{EOFg}.

\subsection{WKB expansion of the wave function}

When $\hbar$ is small but non-zero, we can WKB expand $\psi(x)$, i.e.:
\beq
\psi(x) \sim \ee{-{1\over 2\hbar} \int^x \Ycl(x')dx'}\,\,{1\over \sqrt{\Ycl(x)}}\,\left(1+\sum_k \hbar^k \psi_k(x)\right)
\eeq
i.e.
\beq
Y \sim \Ycl + \sum_{k=1}^\infty \hbar^k \,Y_k
\eeq
The expansion coefficients $Y_k$ can be easily obtained recursively from the Ricatti equation:
\beq
2\Ycl Y_k  = 2Y_{k-1}' - \sum_{j=1}^{k-1} Y_j\,Y_{k-j}
\eeq
For instance:
\beq
Y_1 = {\Ycl'\over \Ycl}
\virg
Y_2 = {Y'_1\over \Ycl} - {Y_1^2\over 2 \Ycl} = {\Ycl''\over \Ycl^2}-{3\over 2}\,{{\Ycl'}^2\over \Ycl^3}
\virg
\dots\, {\rm etc}
\eeq

\subsection{$\hbar$ expansion of correlators and energies}

The kernel $K(x_0,x)$ can also be expanded:
\beq
K(x_0,x) = K_{\rm cl}(x_0,x) + \sum_{k=1}^\infty \hbar^k K_{(k)}(x_0,x)
\eeq
where $K_{(0)}=K_{\rm cl}$ is the kernel of \cite{EOFg}:
\beq
K_{\rm cl}(x_0,x) = {dE_{x,o}(x_0)\over Y_{\rm cl}(x)}
\eeq

This implies that the correlators $W_n^{(g)}$ can also be expanded:
\beq
 W^{(g)}_{n}(x_1,\dots,x_n)  = \sum_{k=0}^\infty \hbar^k\,\, W^{(g,k)}_{n}(x_1,\dots,x_n) 
\eeq
where the $W^{(g,k)}_{n}$ are obtained by the recursion:
\bea
 W^{(g,k)}_{n+1}(x_0,J) 
&=&   \sum_{l=0}^k \sum_i \Res_{x\to s_i}  K_{(k-l)}(x_0,x)\, \Big[ \overline{W}_{n+2}^{(g-1,l)}(x,x,J)  \cr
&& + \sum_{h=0}^g\sum_{j=0}^{l} \sum_{I\subset J}' W_{|I|+1}^{(h,j)}(x,I) W_{n-|I|+1}^{(g-h,l-j)}(x,J/I) \Big] 
\eea
where $J=\{ x_{1},\dots,x_{n} \}$.

Therefore, we observe that to leading order in $\hbar$, the $\lim_{\hbar\to 0} W^{(g,k)}_{n}=W^{(g,0)}_{n}$ do coincide with the $W^{(g)}_{n}$ computed with only $K_{\rm cl}$, and thus they coincide with the $W^{(g)}_{n}$ of \cite{EOFg}.

And also, the $\hbar$ expansion must coincide with  the diagrammatic rules of \cite{ChekEynbeta}.

\section{Non-commutative algebraic geometry}
\label{secqalgeo}

We have seen that in the limit $\hbar\to 0$, the correlation functions and the various functions we are considering, are fundamental objects of algebraic geometry.
For instance $B$ is the Bergmann kernel, and $K$ is the recursion kernel of \cite{EOFg}, which generates the symplectic invariants $F_g$ and the correlators $W_n^{(g)}$ attached to the spectral curve $\Ycl(x)$.

\smallskip

In this paper, when $\hbar\neq 0$, we have defined deformations of those objects, which have almost the same properties as the classical ones, except that they are no longer algebraic functions.

For instance we have:
\begin{itemize}

\item {\bf Spectral curve}

The algebraic equation of the classical spectral curve is replaced by a linear differential equation:
\beq
0={\cal E}(x,y)=\sum_{i,j} {\cal E}_{i,j}\, x^i y^j
\quad \to \quad
0={\cal E}(x,\hbar \partial) \psi=\sum_{i,j} {\cal E}_{i,j}\, x^i (\hbar \partial)^j\,\psi
\eeq
In other words the polynomial ${\cal E}(x,y)$ is replaced by a non-commutative polynomial with $y=\hbar \partial_x$, i.e. $[y,x]=\hbar$.

Here, our non-commutative spectral curve is:
\beq
{\cal E}(x,y) = y^2 - U(x)
\virg y=\hbar \partial_x
\eeq
Notice that it can be factorized as:
\beq
{\cal E}(x,y) = (y-{Y\over 2})\,(y+{Y\over 2})
\eeq
where $Y(x)$ is solution of $Y^2-2\hbar Y'=U$.

\item {\bf Bergmann Kernel} $B(x_1,x_2)$

The non-commutative Bergmann kernel $B(x_1,x_2)$ is closely related to the Inverse of the Hessian $T$, i.e. to $A=T^{-1}$:
\beq
B(x_1,x_2) = {1\over 2(x_1-x_2)^2} + \sum_{i,j} {A_{i,j}\over (x_1-s_i)^2(x_2-s_j)^2}
\eeq
A property of the classical Bergmann kernel $B_{\rm cl}(x_1,x_2)$ is that it computes derivatives, i.e. for any meromorphic function $f(x)$ defined on the spectral curve we have:
\beq
df(x) = -\Res_{x_2\to {\rm poles\, of}\, f}\, B_{\rm cl}(x,x_2)\, f(x_2)
\eeq
Here, this property is replaced by:
for any function $f(x)$ defined on the non-commutative spectral curve (i.e. with poles only at the $s_i$'s), we have:
\beq
f'(x) = -2\sum_i \Res_{x_2\to s_i}\, B(x,x_2)\, f(x_2)\,\, dx_2
\eeq
The factor of $2$, comes from the fact that the interpretation of $x$, and thus of derivatives with respect to $x$, is slightly different.
In the classical case, the differentials are computed in terms of local variables, and $x$ is not a local variable near branch-points. A good local variable near a branchpoint $a$, is $\sqrt{x-a}$.
In the non-commutative case, the role of branchpoints seems to be played by the $s_i$'s, and $x$ is a good local variable near $s_i$.

\item {\bf Rauch variational formula}:
In classical algebraic geometry, on an algebraic curve of equation ${\cal E}(x,y)=\sum_{i,j} {\cal E}_{i,j} x^i y^j=0$, the Bergmann kernel depends only on the location of branchpoints $a_i$. The branchpoints are the points where the tangent is vertical, i.e. $dx(a_i)=0$. Their location is $x_i=x(a_i)$.
The Bergmann kernel is only function of the $x_i$'s, and the classical variational Rauch formula reads:
\beq
{\partial\, B_{\rm cl}(z_1,z_2)\over \partial x_i} = \Res_{z\to a_i}\, {B_{\rm cl}(z,z_1)\,B_{\rm cl}(z,z_2)\over dx(z)}
\eeq
Equivalently, we can parametrize the spectral curve as $x(y)$ instead of $y(x)$, and consider the branchpoints of $y$, i.e. $dy(b_i)=0$, whose location is $y_i=y(b_i)$, and we have:
\beq\label{ClassicalRauch}
{\partial\, B_{\rm cl}(z_1,z_2)\over \partial y_i} = \Res_{z\to b_i}\, {B_{\rm cl}(z,z_1)\,B_{\rm cl}(z,z_2)\over dy(z)}
\eeq

Here, in the non-commutative version, theorem \ref{thW3Krich} and theorem \ref{thvariationV} implies that under a variation of the spectral curve, we have:
\beq
\delta B(x_1,x_2) =-{1\over 2}\,\sum_i \Res_{x\to s_i} {B(x,x_1)B(x,x_2)\over Y'(x)}\,\, \delta Y(x)
\eeq
Consider the branchpoints $b_i$ such that $Y'(b_i)=0$, and define their location as $Y_i=Y(b_i)$, by moving the integration contours we have:
\bea
\delta B(x_1,x_2) 
&=& {1\over 2}\,\sum_i \Res_{x\to b_i} {B(x,x_1)B(x,x_2)\over Y'(x)}\,\, \delta Y(x)\, dx \cr
&=& {1\over 2}\,\sum_i \delta Y_i\, \Res_{x\to b_i} {B(x,x_1)B(x,x_2)\over Y'(x)}\,dx \cr
\eea
i.e.:
\beq
{\partial\, B(x_1,x_2)\over \partial Y_i} = {1\over 2}\,\Res_{x\to b_i}\, {B(x,x_1)\,B(x,x_2)\over Y'(x)}\,\, dx
\eeq
which is thus the quantum version of the Rauch variational formula \eq{ClassicalRauch}.

\end{itemize}

\bigskip
Those properties can be seen as the beginning of a dictionary giving the deformations of classical algebraic geometry into non-commutative algebraic geometry.

\bigskip

{\bf Conjecture about the symplectic invariants}

The $F_g$'s of \cite{EOFg} are the symplectic invariants of the classical spectral curve,
which means that they are invariant under any cannonical change of the spectral curve which conserves the symplectic form $dx\wedge dy$.
For instance they are invariant under $x\to y,y\to -x$.

Here, we conjecture that we may define some non-commutative $F^{(g)}$'s which are invariant under any cannonical transformation which conserves the commutator $[y,x]=\hbar$.
This duality should also correspond to the expected duality $\beta\to 1/\beta$ in matrix models, cf \cite{mkrtchyan1, Bryc}.

However, to check the validity of this conjecture, one needs to extend our work to differential operators of any order in $y$, and not only order $2$. We plan to do this in a forthcoming work.

\section{Application: non-hermitian Matrix models}
\label{secMM}

The initial motivation for the work of \cite{EOFg}, as well as this present work, was initially random matrix models. The classical case corresponds to hermitian matrix models, and here, we show that $\hbar\neq 0$ corresponds in some sense to non-hermitian matrix models \cite{BrezNb, Bryc, Dum}.

\smallskip

In this section, we show that non-hermitian matrix models satisfy the loop equation \eq{loopeqappgen} of theorem \ref{thWngPng}.

We define the matrix integral over $E_{m,2\beta}=$set of $m\times m$ matrices of Wigner--type $2\beta$ ($E_{m,1}=$ real symmetric matrices, $E_{m,2}=$ hermitean matrices, $E_{m,4}=$ real quaternion self-dual matrices, see \cite{Mehtabook}):
\beq
Z = \int_{E_{m,2\beta}} dM\,\, \ee{-{N\sqrt\beta}\, \Tr V(M)}
\eeq
where $N$ is some arbitrary constant, not necessarily related to the matrix size $m$.

It is more convenient to rewrite it in terms of eigenvalues of $M$ (see \cite{Mehtabook}):
\beq
Z = \int_{{\cal C}^m}\, d\l_1\dots d\l_m\,\, \prod_{i>j} (\l_j-\l_i)^{2\beta}\,\, \prod_i \ee{-{N\sqrt\beta}\, V(\l_i)}
\eeq
This last expression is well defined for any $\beta$, and not only $1/2,1,2$, and for any contour of integration ${\cal C}$ on which the integral is convergent.

We also define the correlators:
\bea
\overline{W}_n(x_1,\dots,x_n) 
&=& <\Tr{1\over x_1-M}\dots \Tr {1\over x_n-M}>_c  \cr
&=& \left(N\sqrt\beta\right)^{-n}\,\,{\partial \over \partial V(x_1)}\dots{\partial \over \partial V(x_n)}\,\ln{Z}
\eea
i.e. in terms of eigenvalues:
\beq
\overline{W}_n(x_1,\dots,x_n) = <\sum_{i_1} {1\over x_1-\l_{i_1}} \dots \sum_{i_n} {1\over x_n-\l_{i_n}}>_c
\eeq
In order to match with the notations of section \ref{secdefWngFg}, we prefer to shift $\overline{W}_2$ by a second order pole, and we define:
\beq
W_n(x_1,\dots,x_n) = \overline{W}_n(x_1,\dots,x_n) + {\delta_{n,2}\over 2(x_1-x_2)^2}
\eeq

\bigskip

We are interested in a case where $Z$ has a large $N$ expansion of the form:
\beq
\ln{Z} \sim \sum_{g=0}^\infty N^{2-2g}\,\, F_g 
\eeq
and for the correlation functions we assume:
\beq\label{Wntopoexp}
W_{n}( x_1,\dots,x_n) = {1\over \beta^{n/2}}\,\,\sum_{g=0}^\infty N^{2-2g-n} W_{n}^{(g)}(x_1,\dots,x_n)
\eeq

\subsection{Loop equations}

The loop equations can be obtained by integration by parts, or equivalently, they follow from the invariance of an integral under a change of variable.
By considering the infinitesimal change of variable:
\beq
\l_i \to \l_i + \epsilon{1\over x-\l_i} + O(\epsilon^2)
\eeq
we obtain:
\bea\label{recMMbetaWallkg}
&& {N\sqrt\beta}(V'(x)\, \overline{W}_{n+1}(x, x_1,\dots,x_n) - P_{n+1}(x; x_1,\dots,x_n) ) \cr
&=&  \beta \sum_{J\subset L}  \overline{W}_{1+|J|}(x,J)\,\overline{W}_{1+n-|J|}(x,{L/J})  \cr
&& +  \beta \overline{W}_{n+2}(x,x, x_1,\dots,x_n) \cr
&& - (1-\beta)\,{\partial \over \partial x} \, \overline{W}_{n+1}(x, x_1,\dots,x_n) \cr
&& + \sum_{j=1}^n  {\partial \over \partial x_j} \, {\overline{W}_{n}(x, L/\{x_j\})-\overline{W}_{n}(x_j, L/\{x_j\})\over x-x_j}
\eea
where $P_{n+1}(x; x_1,\dots,x_n) )$ is a polynomial in its first variable $x$, of degree $\delta_{n,1}+\deg V-2$.

If we expand this equation into powers of $N$ using \eq{Wntopoexp}, we have $\forall\, n,g$:
\bea\label{recMMbetaWg}
&& V'(x)\, \overline{W}_{n+1}^{(g)}(x, x_1,\dots,x_n) - P_{n+1}^{(g)}(x; x_1,\dots,x_n) ) \cr
&=&  \sum_{g'=0}^g \sum_{J\subset L}  \overline{W}_{1+|J|}^{(g')}(x,J)\,\overline{W}_{1+n-|J|}^{(g-g')}(x,{L/J})  \cr
&& +  \beta \overline{W}_{n+2}^{(g-1)}(x,x, x_1,\dots,x_n) \cr
&& +\hbar \,{\partial \over \partial x} \, \overline{W}^{(g)}_{n+1}(x, x_1,\dots,x_n) \cr
&& + \sum_{j=1}^n  {\partial \over \partial x_j} \, {\overline{W}^{(g)}_{n}(x, L/\{x_j\})-\overline{W}^{(g)}_{n}(x_j, L/\{x_j\})\over x-x_j}
\eea
where
\beq
\hbar = {\sqrt\beta-{1\over \sqrt\beta}\over N}
\eeq

Those loop equations coincide with 
the loop equations \eq{loopeqPng} of theorem \ref{thWngPng}.

Moreover we have:
\beq\label{dWngdV}
\overline{W}^{(g)}_n = {\partial \overline{W}_{n-1}^{(g)}\over \partial V}
\eeq
and near $x\to \infty$:
\beq
\sqrt\beta\,\,W_1(x) \sim {m\over x} [N\hbar-\sum_{g=1}^\infty (-1)^g {(2g-2)!\over g! (g-1)!}\,\, (N\hbar)^{1-2g}   ]
\eeq
i.e.
\beq
W^{(0)}_1(x) \sim  {m\hbar \over x} +O(1/x^2)
\virg
W^{(g)}_1(x) \sim - {m\hbar \over x} \,\, \hbar^{-2g}\,\, {(2g-2)!\over g! (g-1)!}+O(1/x^2)
\eeq

One should notice that the loop equations are independent of the contour ${\cal C}$ of integration of eigenvalues.
The contour ${\cal C}$ is in fact encoded in the polynomial $P_{n+1}(x; x_1,\dots,x_n)$.

\subsection{Solution of loop equations}

To order $g=0, n=1$ we have:
\beq
 V'(x)\, W_{1}^{(0)}(x) - P_{1}^{(0)}(x)
= W_{1}^{(0)}(x)^2 +\hbar \,{\partial \over \partial x} \, W_{1}^{(0)}(x)
\eeq
which is the same as the Ricatti equation \eq{eqRicattiUY}.

As we said above, the contour ${\cal C}$ is in fact encoded in the polynomial $P^{(0)}_{1}(x) $.
From now on, we choose a contour ${\cal C}$, i.e. a polynomial $P_{1}^{(0)}(x)$ such that the solution of the Ricatti equation is rational:
\beq
W_{1}^{(0)}(x) = \hbar \sum_{i=1}^m {1\over x-s_i}
\eeq
It also has the correct behaviour at $\infty$: $W_{1}^{(0)}(x)\sim {m\hbar\over x}$.
This corresponds to a certain contour ${\cal C}$ which we do not determine here.

Since $W_1^{(0)}(x) =\om(x)$ satisfies the Ricatti equation, i.e. the Bethe ansatz, the kernel $K$ exists, and we can define the functions $K(x_0,x)$, $G(x_0,x)$ and $B(x_0,x)$.

\bigskip

Then, from \eq{dWngdV}, we see that every $\overline{W}_n^{(g)}$ is going to be a rational fraction of $x$, with poles only at the $s_i$'s.
In particular, Cauchy theorem implies:
\beq
\overline{W}_{n+1}^{(g)}(x_0,x_1,\dots,x_n) = \Res_{x\to x_0} G(x_0,x)\,\overline{W}_{n+1}^{(g)}(x,x_1,\dots,x_n)
\eeq
and since both $G(x_0,x)$ and $\overline{W}_{n+1}^{(g)}(x,x_1,\dots,x_n)$ are rational fractions, which vanish sufficientely at $\infty$, we may change the integration contour to the other poles of the integrand, namely:
\bea
&& \overline{W}_{n+1}^{(g)}(x_0,x_1,\dots,x_n) \cr
&=& -\sum_{i} \Res_{x\to s_i} G(x_0,x)\,\overline{W}_{n+1}^{(g)}(x,x_1,\dots,x_n) \cr
&=& -\sum_{i} \Res_{x\to s_i} \overline{W}_{n+1}^{(g)}(x,x_1,\dots,x_n)\,\, (2\om(x)-V'(x)-\hbar\partial_x)K(x_0,x) \cr
&=& -\sum_{i} \Res_{x\to s_i} K(x_0,x)\,\, (2\om(x)-V'(x)+\hbar\partial_x)\overline{W}_{n+1}^{(g)}(x,x_1,\dots,x_n) \cr
\eea

Now, we insert loop equation \eq{recMMbetaWg} in the right hand side, and we notice that the term 
$P_{n+1}^{(g)}$ and ${\partial \over \partial x_j} \, {W^{(g)}_{n}(x_j, L/\{x_j\})\over x-x_j}$ do not have poles at the $s_i$'s, so they don't contribute. We thus get:
\bea
&& \overline{W}_{n+1}^{(g)}(x_0,x_1,\dots,x_n) \cr
&=& \sum_{i} \Res_{x\to s_i}  K(x_0,x)\, \Big(
\overline{W}_{n+2}^{(g-1)}(x,x,x_1,\dots,x_n) \cr
&& + \sum_{g'=0}^g \sum_{J\subset L} W_{1+|J|}^{(g')}(x,J) W_{1+n-|J|}^{(g-g')}(x,L/J) \Big)
\eea
i.e. we find the correlators of def \ref{defWng}.

Special care is needed for $W_2^{(0)}$.
We have:
\bea
&& \overline{W}_{2}^{(0)}(x_0,x_1,\dots,x_n) \cr
&=& -\sum_{i} \Res_{x\to s_i} K(x_0,x)\,\, (2\om(x)-V'(x)+\hbar\partial_x)\overline{W}_{2}^{(0)}(x,x_1) \cr
&=& \sum_{i} \Res_{x\to s_i} K(x_0,x)\,\, {\om(x)\over (x-x_1)^2} \cr
&=& \hbar\,\sum_{i}  {K(x_0,s_i)\over (s_i-x_1)^2} \cr
&=&\sum_{i,j}  {A_{i,j}\over (s_i-x_1)^2(s_j-x_0)^2} \cr
\eea
which also agrees with  def \ref{defWng}.

\section{Application: Gaudin model}
\label{secGaudin}

The Gaudin model's Bethe ansatz is obtained for the potential:
\beq
V_{\rm Gaudin}'(x) = x+ \sum_{i=1}^{\npole} {S_i\over x-\alpha_i}
\eeq
i.e. it corresponds to a Gaussian matrix model with sources:
\beq
Z= \int_{E_{m,2\beta}} dM\,\, \ee{-{N\sqrt\beta \over 2}\Tr M^2}\,\, \prod_i \det(\alpha_i-M)^{-N S_i \sqrt\beta}
\eeq
with $\hbar = {\sqrt\beta-1/\sqrt\beta\over N}$.

$Z$ can also be written in eigenvalues:
\beq
Z = \int d\l_1\dots d\l_m \,\, {\prod_{i=1}^m \ee{-{N\sqrt\beta\over 2}\l_i^2}\over \prod_{i=1}^m\prod_{j=1}^{\ovl{n}} (\alpha_j-\l_i)^{N\sqrt\beta\, S_j}}\,\,\, \prod_{i>j} (\l_i-\l_j)^{2\beta}
\eeq

\subsection{Example}

Consider:
\beq
V'(x) = x-{s^2\over x}
\virg
V(x) = {x^2\over 2} - s^2\ln{x}
\eeq
With only 1 root $m=1$, the solution of the Bethe equation $V'(x)=0$ is $x=s$.

Thus we have:
\beq
\om(x) = {\hbar\over x-s}
\eeq
\beq
B(x_1,x_2)={1\over 2(x_1-x_2)^2} + {\hbar\over 2(x_1-s)^2(x_2-s)^2}
\eeq

We find:
\beq
W_3^{(0)}(x_1,x_2,x_3) 
= {\hbar\over 2(x_1-s)^2(x_2-s)^2(x_3-s)^2}\,\left({1\over x_1-s}+{1\over x_2-s}+{1\over x_3-s}+{1\over 2s}\right) 
\eeq
\beq
W_1^{(1)}(x) = {1\over \hbar(x-s)} +{1\over 4s(x-s)^2}+{1\over 2(x-s)^3}
\eeq

For the free energies we have:
\beq
F^{(0)} = {\hbar\,s^2\over 2}\,(\ln{s^2}-1)
\eeq
\beq
F^{(1)} = {1\over 2}\,\ln{({\hbar\over 2})} + {F^{(0)}\over \hbar^2}
\eeq
\beq
F^{(2)} = - {1\over 12 \hbar s^2} -  {F^{(0)}\over \hbar^4}
\eeq
\beq
F^{(3)} =  {1\over 12 \hbar^3 s^2} + {2F^{(0)}\over \hbar^6}
\eeq
and
\beq
Z = \ee{\sum_g N^{2-2g} F^{(g)}} = \ee{-N\sqrt\beta V(s)}\,{1\over \sqrt{2\hbar}}\,\,(1-{1\over 12 s^2 N^2 \hbar^2} + \dots)
\eeq
which is indeed the beginning of the saddle point expansion of:
\beq
Z =  \int dx\,\, \ee{-N\sqrt\beta\,\, V(x)}
\eeq

\section{Conclusion}
\label{secConcl}

In this article, we have defined a special case of non-commutative deformation of the symplectic invariants of \cite{EOFg}. Many of the fundamental properties of \cite{EOFg} are conserved or only slightly modified.

The main difference, is that the recursion kernel, instead of beeing an algebraic function, is given by the solution of a differential equation, otherwise the recursion is the same.

\bigskip

The main drawback of our definition, is that it concerns only a very restrictive subset of possible non-commutative spectral curves.
Namely, we considered here only non commutative polynomials ${\cal E}(x,y)=\sum_{i,j} {\cal E}_{i,j}\,\, x^i y^j$ with $y=\hbar \partial_x$, of degree 2 in $y$, and such that the differential equation ${\cal E}(x,\hbar\partial).\psi=0$ has a "polynomial" solution of the form $\psi(x)=\prod_{i=1}^m (x-s_i)\,\, \ee{-V(x)/2\hbar}$.

It should be possible to extend our definitions to other "non-polynomial" solutions $\psi$ (with an infinite number of zeroes $m=\infty$ for instance), and/or to higher degrees in $y$.
In other words, what we have so far, is only a glimpse on more general structure yet to be discovered.

For example, it is not yet clear how our definitions are related to matrix integrals.
We have said that the integration contour for the eigenvalues should be chosen so that the solution of the Schroedinger equation is polynomial of degree $m$, however, it is not known how to find explicitly such integration contours. Conversely, the usual matrix integrals with eigenvalues on the real axis, do probably not correspond to polynomial solutions of the Schroedinger equation.
Similarly, it is not clear what the relationship between our definitions and the number of unoriented ribbon graphs is, for the same reason. The solution of the Schroedinger equation for ribbon graphs, should be chosen such that all the $W_n^{(g,k)}$'s are power series in $t$, and it is not known which integration contour it corresponds to, and which solution of the Schroedinger equation it corresponds to.

\bigskip

Therefore it seems necessary to extend our definitions to arbitrary solutions, i.e. to arbitrary integration contours for the matrix integrals. A possibility could be to obtain non-polynomial solutions as limits of polynomial ones.

\medskip

The extension to higher degree in $y$, can be obtained from multi-matrix integrals, and extension seems rather easy for polynomial solutions again.

\medskip

Finally, like the symplectic invariants of \cite{EOFg}, we expect those "to be defined" non-commutative symplectic invariants, to play a role in several applications to enumerative geometry, and to topological string theory like in \cite{remodelBmodel}. In other words, we expect our $F^{(g)}$'s to be generating functions for intersection numbers in some non-commutative moduli spaces of unoriented Riemann surfaces, whatever it means...

\section*{Acknowledgments}
We would like to thank O. Babelon, M. Berg\`ere, M. Bertola, L. Chekhov, R. Dijkgraaf, J. Harnad and N. Orantin for useful and fruitful discussions on this subject.
This work is partly supported by the Enigma European network MRT-CT-2004-5652, by the ANR project G\'eom\'etrie et int\'egrabilit\'e en physique math\'ematique ANR-05-BLAN-0029-01, by the Enrage European network MRTN-CT-2004-005616,
by the European Science Foundation through the Misgam program,
by the French and Japaneese governments through PAI Sakurav, by the Quebec government with the FQRNT.


\setcounter{section}{0}

\appendix{Expansion of $K$}\label{appKexpansion}

Since we have to compute residues at the $s_i$'s, we need to compute the Taylor expansion of $K(x_0,x)$ when $x\to s_i$:
\beq\label{TaylexpKikx0}
K(x_0,x) = \sum_{k} (x-s_i)^k\,K_{i,k}(x_0)
\eeq
For instance we find:
\beq
K_{i,0} = {1\over \hbar}\sum_{j}\, {A_{i,j}\over (x_0-s_j)^2}
\eeq
\bea
 \hbar K_{i,1} (x_0)
&=& 
-{1\over (x_0-s_i)} 
- 2\sum_{a\neq i} \sum_{j} {A_{a,j}\over (s_a-s_i)\,(x_0-s_j)^2}   \cr
\eea
\bea
 \hbar K_{i,3} 
&=& - \hbar\Big(   2\sum_{a\neq i} {1\over (s_a-s_i)^{2}} + {1\over \hbar} V''(s_i) \Big)\, K_{i,1} \cr
&& - \hbar\Big(   2\sum_{a\neq i} {1\over (s_a-s_i)^{3}} + {1\over \hbar}  {V'''(s_i)\over 2}  \Big)\, K_{i,0}  \cr
&& +{1\over (x_0-s_i)^{3}} 
+ 2\sum_{a\neq i} \sum_{j} {A_{a,j}\over (s_a-s_i)^{3}\,(x_0-s_j)^2}   \cr
\eea
Thanks to property \eq{Ki2vanishhyp}, we may assume (but it is not necessary) that:
\beq
K_{i,2}=0
\eeq
Then, we have the recursion for $k\geq 0$:
\bea
&& \hbar\Big( (1-k)K_{i,k+1} - 2\sum_{a\neq i}\sum_{l=0}^{k} {K_{i,k-l}\over (s_a-s_i)^{l+1}} - {1\over \hbar} \sum_{l=0}^k {V^{(l+1)}(s_i)\over l!} K_{i,k-l} \Big) \cr
&=& -{1\over (x_0-s_i)^{k+1}} 
- 2\sum_{a\neq i} \sum_{j} {A_{a,j}\over (s_a-s_i)^{k+1}\,(x_0-s_j)^2}  
\eea
This proves that each $K_{i,k}(x_0)$ is a rational fraction of $x_0$, with poles at the $s_j$'s.

\subsection{Rational fraction of $x_0$}

Thus we write:
\beq\label{TaylexpKikjl}
K_{i,k}(x_0) = \sum_{j,l} {1\over (x_0-s_j)^{k'}}\,\, K_{i,k;j,k'} 
\eeq
For instance we have:
\beq
K_{i,0;j,k'} = {A_{i,j}\over \hbar}\,\delta_{k',2}
\eeq
\beq
 \hbar K_{i,1;j,k'} = 
 - \delta_{k',1}  \delta_{i,j}
- 2 \delta_{k',2}\, \sum_{a\neq i} {A_{a,j}\over s_a-s_i}   
\eeq
For higher $k$ we have the recursion:
\bea
&& \hbar\Big( (1-k)K_{i,k+1;j,k'} - 2\sum_{a\neq i}\sum_{l=1}^{k} {K_{i,k-l;j,k'}\over (s_a-s_i)^{l+1}} - {1\over \hbar} \sum_{l=1}^k {V^{(l+1)}(s_i)\over l!} K_{i,k-l;j,k'} \Big) \cr
&=& - \delta_{i,j}\, \delta_{k',k+1} 
- 2\delta_{k',2}\,\sum_{a\neq i} {A_{a,j}\over (s_a-s_i)^{k+1}}  
\eea
In particular, it shows that if $k'>2$, then $K_{i,k;i,k'}$ is proportional to $\delta_{i,j}$.

\subsection{Generating functions}

We introduce generating functions:
\beq
R_{i;j,k'}(x) = \sum_i K_{i,k;j,k'}\, (x-s_i)^k
\eeq
We have:
\bea
 \hbar\,\Big(  2{\psi'(x)\over \psi(x)} -\partial_x \Big)\, R_{i;j,k'}(x)  
= - \delta_{i,j} (x-s_i)^{k'-1} 
+ 2\delta_{k',2}\,\sum_{a} {A_{a,j}\over x-s_a}  
\eea
i.e.
\bea
- \hbar \psi^2(x)\,\partial_x\Big(  { R_{i;j,k'}(x)  \over \psi^2(x)}\Big)
= - \delta_{i,j} (x-s_i)^{k'-1} + \delta_{k',1} c_j
+ 2\delta_{k',2}\,\sum_{a} {A_{a,j}\over x-s_a}  
\eea

In particular with $k'=1$ we find:
\beq\label{Rij1x}
R_{i;j,1}(x) = 
{\delta_{i,j}\over \hbar}\, \psi(x) \phi(x)
\eeq
where
\beq
\phi(x) = \psi(x)\int^x {dx'\over \psi(x')^2}
\virg
\phi'(x)\psi(x) - \psi'(x)\phi(x) = 1
\eeq

\appendix{Proof of theorem \ref{thpolessiWng}}
\label{approofthpolessiWng}

{\bf Theorem \ref{thpolessiWng}}
{\it Each $W_n^{(g)}$ is a rational function of all its arguments. If $2g+n-2>0$, it has poles only at the $s_i$'s. In particular it has no poles at the $\alpha_i$'s, and it vanishes as $O(1/x_i)$ when $x_i\to\infty$.
}
\bigskip

{\bf proof:}

It is easy to check that $W_1^{(0)}$, $W_2^{(0)}$ satisfy the theorem. 

We will now make a recursion over $-\chi=2g-2+n$ to prove the result for every $(n,g)$.
We write:
\beq 
W^{(g)}_{n+1}(x_0,x_1,\dots,x_n)
=   \sum_i \Res_{x\to s_i}  K(x_0,x)\,  U^{(g)}_{n+1}(x,x_1,\dots,x_n)
\eeq
where $J=\{ x_1,\dots,x_n\}$, and
\beq\label{defUng1}
U^{(g)}_{n+1}(x,J)
=    \ovl{W}_{n+2}^{(g-1)}(x,x,J) + \sum_{h=0}^g \sum_{I\subset J} W_{|I|+1}^{(h)}(x,I) W_{n-|I|+1}^{(g-h)}(x,J/I)  
\eeq
First, the recursion hypothesis clearly implies that $U_{n+1}^{(g)}(x,x_1,\dots,x_n)$ is a rational fraction in all its variables $x,x_1,...x_n$.

Then we Taylor expand $K(x_0,x)$ as in \eq{TaylexpKikx0} or \eq{TaylexpKikjl}
\bea 
W^{(g)}_{n+1}(x_0,x_1,\dots,x_n)
&=&   \sum_i \Res_{x\to s_i}  K(x_0,x)\,  U^{(g)}_{n+1}(x,x_1,\dots,x_n) \cr
&=&   \sum_i \sum_k K_{i,k}(x_0)\,\, \Res_{x\to s_i}  (x-s_i)^k\,  U^{(g)}_{n+1}(x,x_1,\dots,x_n) \cr
\eea
Since $U_{n+1}^{(g)}(x,x_1,\dots,x_n)$ is a rational fraction of $x$, the sum over $k$ is finite, and
therefore, $W^{(g)}_{n+1}(x_0,x_1,\dots,x_n)$ is a finite sum of rational fractions of $x_0$, with poles at the $s_j$'s, therefore it is a rational fraction of $x_0$ with poles at the $s_j$'s.

It is also clear that $W^{(g)}_{n+1}(x_0,x_1,\dots,x_n)$ is a rational fraction of the other variables $x_1,\dots,x_n$.
 The poles in those variables are necessarily at the $s_j$'s, because as long as the residues can be computed, $W^{(g)}_{n+1}(x_0,x_1,\dots,x_n)$ is finite. The residue cannot be computed everytime an integration contour gets pinched, and since the integration contours are small circles around the $s_i$'s, the only singularities may occur at the $s_i$'s.

\medskip
It remains to prove that each $W_n^{(g)}$ behaves like $O(1/x_i)$ at $\infty$. The proof follows the same line: each $K_{i,k}(x_0)$ behaves like $O(1/x_0)$, and by an easy recursion the result holds for all other variables.
$\square$

\appendix{Proof of theorem \ref{thWngPng}}
\label{approofthWngPng}

In this subsection we prove theorem \ref{thWngPng}, that all $W_n^{(g)}$'s satisfy
the loop equation.

\medskip
{\bf Theorem \ref{thWngPng}}
{\it 
 The $W_n^{(g)}$'s satisfy the loop equation, i.e. the following quantity $P_{n+1}^{(g)}(x;x_1...,x_n)$
\bea\label{loopeqappgen}
 P_{n+1}^{(g)}(x;x_1...,x_n)
 &=&
-Y(x)\overline{W}_{n+1}^{(g)}(x,x_1,...,x_n) + \hbar \partial_{x}{\overline{W}_{n+1}^{(g)}(x,x_1...,x_n)} \cr
&& + \sum_{I\subset J} \ovl{W}_{|I|+1}^{(h)}(x,x_I) \ovl{W}_{n-|I|+1}^{(g-h)}(x,J/I) +
\ovl{W}_{n+2}^{(g-1)}(x,x,J)  \cr
& &+ \sum_{j}
\partial_{x_j} \left( {{\ovl{W}_n^{(g)}(x,J/\{j\})-{\ovl{W}_n^{(g)}(x_j,J/\{j\})}} \over {(x-x_j)}}\right) \cr
\eea
is a rational fraction of $x$ (possibly a polynomial), with no pole at $x=s_i$.
The only possible poles of $P_{n+1}^{(g)}(x;x_1...,x_n)$ are at the poles of $V'(x)$, and their degree is less than the degree of $V'$.
}
\bigskip

{\bf proof:}

First, from theorem \ref{thpolessiWng}, we easily see that $P_{n+1}^{(g)}(x;x_1...,x_n)$ is indeed a rational function of $x$. 
Moreover it clearly has no pole at coinciding points $x=x_j$.

Then we write Cauchy's theorem for $W_{n+1}^{(g)}$:
\bea
W_{n+1}^{(g)}(x_0,...,x_n)
&=&\Res_{x\to x_0} {1 \over{x-x_0}}\, W_{n+1}^{(g)}(x,x_1,...,x_n)  \cr
&=&\Res_{x\to x_0}  G(x_0,x)\,W_{n+1}^{(g)}(x,x_1,...,x_n) 
\eea
and using again theorem \ref{thpolessiWng}, i.e. that $W_{n+1}^{(g)}$ has poles only at the $s_i$'s, and that both $W_{n+1}^{(g)}$ and $G(x_0,x)$ behave as $O(1/x)$ for large $x$, we may move the integration contours:
\beq
W_{n+1}^{(g)}(x_0,...,x_n)
= -\sum_i \Res_{x\to s_i} G(x_0,x)\,\, W_{n+1}^{(g)}(x,x_1,...,x_n) 
\eeq
Then we use the definition of $K$, and integrate by parts:
 \bea
 W_{n+1}^{(g)}(x_0,...,x_n)   
&=&\sum_i \Res_{x\to s_i} (Y(x)K(x_0,x)+ \hbar K'(x_0,x))W_{n+1}^{(g)}(x,x_1,...,x_n)  \cr
&=& \sum_i \Res_{x\to s_i} K(x_0,x)\,
\Big(Y(x)W_{n+1}^{(g)}(x,x_1,...,x_n)\cr
&-& \hbar \partial_x
W_{n+1}^{(g)}(x,x_1,...,x_n) \Big) \cr
\eea 
From the definition we have also 
\bea 
&& W_{n+1}^{(g)}(x_0,...,x_n) \cr &=& \sum_i
\Res_{x\to s_i} K(x_0,x) \left(\sum_{h=0}^g \sum_{I\subset J}
W_{|I|+1}^{(h)}(x,I) W_{n-|I|+1}^{(g-h)}(x,J/I)
+\ovl{W}_{n+2}^{(g-1)}(x,x,J)\right) \cr
\eea 
then we shift $W_n^{(g)}$ to $\ovl{W}_n^{(g)}$ in the RHS, i.e.: 
\bea &&
W_{n+1}^{(g)}(x_0,...,x_n) \cr &=& \sum_i \Res_{x\to s_i} K(x_0,x)
\Big(\sum_{h=0}^g \sum_{I\subset J} \ovl{W}_{|I|+1}^{(h)}(x,I)
\ovl{W}_{n-|I|+1}^{(g-h)}(x,J/I) + \ovl{W}_{n+2}^{(g-1)}(x,x,J) \cr
&& + \sum_{j=1}^n  {\ovl{W}_n^{(g)}(x,J/\{j\}) \over (x-x_j)^2}
\Big) \cr &=& \sum_i \Res_{x\to s_i} K(x_0,x) \Big(\sum_{h=0}^g
\sum_{I\subset J} \ovl{W}_{|I|+1}^{(h)}(x,I)
\ovl{W}_{n-|I|+1}^{(g-h)}(x,J/I) + \ovl{W}_{n+2}^{(g-1)}(x,x,J) \cr
&& + \sum_{j=1}^n  \partial_{x_j}\,
\left({\ovl{W}_n^{(g)}(x,J/\{j\}) \over x-x_j}\right) \Big) \cr &=&
\sum_i \Res_{x\to s_i} K(x_0,x) \Big(\sum_{h=0}^g \sum_{I\subset J}
\ovl{W}_{|I|+1}^{(h)}(x,I) \ovl{W}_{n-|I|+1}^{(g-h)}(x,J/I) +
\ovl{W}_{n+2}^{(g-1)}(x,x,J) \cr && + \sum_{j=1}^n  \partial_{x_j}\,
\left({\ovl{W}_n^{(g)}(x,J/\{j\})-\ovl{W}_n^{(g)}(x_j,J/\{j\}) \over
x-x_j}\right) \Big) \cr 
\eea 
in the last line we have added for free, the term $\ovl{W}_n^{(g)}(x_j,J/\{j\})$ because it has no pole at $x=s_i$.

Therefore we have:
\bea
0 &=& \sum_i \Res_{x\to s_i} K(x_0,x)
\Big( -Y(x)W_{n+1}^{(g)}(x,x_1,...,x_n)+ \hbar \partial_x W_{n+1}^{(g)}(x,x_1,...,x_n) \cr
&& + \sum_{h=0}^g \sum_{I\subset J} \ovl{W}_{|I|+1}^{(h)}(x,I)
\ovl{W}_{n-|I|+1}^{(g-h)}(x,J/I) + \ovl{W}_{n+2}^{(g-1)}(x,x,J) \cr
&& + \sum_{j=1}^n  \partial_{x_j}\, \left({\ovl{W}_n^{(g)}(x,J/\{j\})-\ovl{W}_n^{(g)}(x_j,J/\{j\}) \over x-x_j}\right) \Big) \cr
&=& \sum_i \Res_{x\to s_i} K(x_0,x) P_{n+1}^{(g)}(x;x_1,...,x_n) \cr
&=& \sum_i \sum_k K_{i,k}(x_0) \,\, \Res_{x\to s_i} (x-s_i)^k P_{n+1}^{(g)}(x;x_1,...,x_n) \cr
\eea
Notice that this equation holds for any $x_0$.
Since $K_{i,k}(x_0)$ is a rational fraction with a pole of degree $k+1$ in $x_0=s_i$, the $K_{i,k}(x_0)$ are linearly independent functions, and thus we must have:
\beq
\forall k,i \qquad 
0=\Res_{x\to s_i}  (x-s_i)^k\, P_{n+1}^{(g)}(x;x_1,...,x_n) 
\eeq
this means that $P_{n+1}^{(g)}$ has no pole at $x=s_i$.

One easily sees that $P_{n+1}^{(g)}(x;x_1,\dots,x_n)$ is a rational fraction of $x$, and its poles are at most those of $Y(x)$, i.e. at the poles of $V'(x)$.
$\square$

\appendix{Proof of theorem \ref{thsym}}
\label{approofthsym}

{\bf Theorem \ref{thsym}}
{\it 
Each $W_n^{(g)}$ is a symmetric function of all its arguments.
}
\bigskip

{\bf proof:}

The special case of $W_3^{(0)}$ is proved in appendix \ref{approofthW3Krich} above.
It is obvious from the definition that $W_{n+1}^{(g)}(x_0,x_1,\dots,x_n)$ is
symmetric in $x_1,x_2,\dots,x_n$, and therefore we need to show that (for $n\geq 1$):
\beq
W_{n+1}^{(g)}(x_0,x_1,J)-W_{n+1}^{(g)}(x_1,x_0,J)=0
\eeq
where $J=\{ x_2,\dots,x_{n}\}$.
We prove it by recursion on $-\chi=2g-2+n$. 

Assume that every $W_k^{(h)}$ with $2h+k-2\leq 2g+n$ is symmetric.
We have:
\bea
&& W_{n+1}^{(g)}(x_0,x_1,J) \cr
&=&\sum_i \Res_{x\to s_i} K(x_0,x)\,\, \Big(
W_{n+2}^{(g-1)}(x,x,x_1,J) + 2 \,\,\, B(x,x_1) W_{n}^{(g)}(x,J) \cr
&& + 2 \sum_{h=0}^g\sum'_{I\in J}\,\,\, W_{2+|I|}^{(h)}(x,x_1,I) W_{n-|I|}^{(g-h)}(x,J/I) \Big) \cr
\eea
where $\sum'$ means that we exclude the terms $(I=\emptyset, h=0)$ and $(I=J, h=g)$. Notice also that $\ovl{W}_{n+2}^{(g-1)}=W_{n+2}^{(g-1)}$ because $n\geq 1$.
Then, using the recursion hypothesis, we have:
\bea
&& W_{n+1}^{(g)}(x_0,x_1,J) \cr
&=& 2 \sum_i \Res_{x\to s_i} K(x_0,x)\,\,  B(x,x_1) W_{n}^{(g)}(x,J) \cr
&& + \sum_{i,j} \Res_{x\to s_i} \Res_{x'\to s_j}\,\, K(x_0,x) K(x_1,x')\,\, 
\Big(  W_{n+3}^{(g-2)}(x,x,x',x',J) \cr
&& + 2\sum_h\sum'_{I} W_{2+|I|}^{(h)}(x',x,I) W_{1+n-|I|}^{(g-1-h)}(x',x,J/I) \cr
&& + 2 \sum_h\sum'_{I} W_{3+|I|}^{(h)}(x',x,x,I) W_{n-|I|}^{(g-1-h)}(x',J/I) \cr
&& + 2 \sum_{h}\sum'_{I\in J}\,\,\, W_{n-|I|}^{(g-h)}(x,J/I)
\Big[ W_{3+|I|}^{(h-1)}(x,x',x',I)  \cr
&& + 2 \sum_{h'}\sum'_{I'\subset I} W_{2+|I'|}^{(h')}(x',x,I')   W_{1+|I|-|I'|}^{(h-h')}(x',I/I')  
\Big]\,
\Big) \cr
\eea
Now, if we compute $W_{n+1}^{(g)}(x_1,x_0,J)$, we get the same expression, with the order of integrations exchanged, i.e. we have to integrate $x'$ before integrating $x$.
Notice, by moving the integration contours,  that:
\beq
\Res_{x\to s_i} \Res_{x'\to s_j} - \Res_{x'\to s_j} \Res_{x\to s_i} =
- \delta_{i,j}\Res_{x\to s_i} \Res_{x'\to x} 
\eeq
Moreover, the only terms which have a pole at $x=x'$ are those containing $B(x,x')$.
Therefore:
\bea
&& W_{n+1}^{(g)}(x_0,x_1,J)-W_{n+1}^{(g)}(x_1,x_0,J) \cr
&=& 2 \sum_i \Res_{x\to s_i} \left( K(x_0,x)\,\,  B(x,x_1)  - K(x_1,x)\,\,  B(x,x_0) \right) \, W_{n}^{(g)}(x,J) \cr
&& - 2 \sum_{i} \Res_{x\to s_i} \Res_{x'\to x}\,\, K(x_0,x) K(x_1,x')\,\, B(x,x')\,\,
\Big(     \cr
&& 2W_{1+n}^{(g-1)}(x',x,J)  + 2 \sum_{h}\sum'_{I\in J}\,\, W_{n-|I|}^{(g-h)}(x,J/I)     W_{1+|I|}^{(h)}(x',I)  
\Big) \cr
\eea
The residue $\Res_{x'\to x}$ can be computed:
\bea
&& W_{n+1}^{(g)}(x_0,x_1,J)-W_{n+1}^{(g)}(x_1,x_0,J) \cr
&=& 2 \sum_i \Res_{x\to s_i} \left( K(x_0,x)\,\,  B(x,x_1)  - K(x_1,x)\,\,  B(x,x_0) \right) \, W_{n}^{(g)}(x,J) \cr
&& -  \sum_{i} \Res_{x\to s_i} \,\, K(x_0,x) {\partial \over \partial x'}\Big( K(x_1,x')\,\, 
\Big(     \cr
&& 2W_{1+n}^{(g-1)}(x',x,J)  + 2 \sum_{h}\sum'_{I\in J}\,\, W_{n-|I|}^{(g-h)}(x,J/I)     W_{1+|I|}^{(h)}(x',I)  
\Big)\,\, \Big)_{x'=x} \cr
&=& 2 \sum_i \Res_{x\to s_i} \left( K(x_0,x)\,\,  B(x,x_1)  - K(x_1,x)\,\,  B(x,x_0) \right) \, W_{n}^{(g)}(x,J) \cr
&& -  \sum_{i} \Res_{x\to s_i}\,\, K(x_0,x) K'(x_1,x)\,\, 
\Big(     \cr
&& 2W_{1+n}^{(g-1)}(x,x,J)  + 2 \sum_{h}\sum'_{I\in J}\,\, W_{n-|I|}^{(g-h)}(x,J/I)     W_{1+|I|}^{(h)}(x,I)  
\Big)\,\,  \cr
&& -  \sum_{i} \Res_{x\to s_i} \,\, K(x_0,x) K(x_1,x) {\partial \over \partial x'}\Big(      \cr
&& 2W_{1+n}^{(g-1)}(x',x,J)  + 2 \sum_{h}\sum'_{I\in J}\,\, W_{n-|I|}^{(g-h)}(x,J/I)     W_{1+|I|}^{(h)}(x',I)  
\,\, \Big)_{x'=x} \cr
&=& 2 \sum_i \Res_{x\to s_i} \left( K(x_0,x)\,\,  B(x,x_1)  - K(x_1,x)\,\,  B(x,x_0) \right) \, W_{n}^{(g)}(x,J) \cr
&& -  \sum_{i} \Res_{x\to s_i} \,\, K(x_0,x) K'(x_1,x)\,\, 
\Big(     \cr
&& 2W_{1+n}^{(g-1)}(x,x,J)  + 2 \sum_{h}\sum'_{I\in J}\,\, W_{n-|I|}^{(g-h)}(x,J/I)     W_{1+|I|}^{(h)}(x,I)  
\Big)\,\,  \cr
&& -  {1\over 2} \sum_{i} \Res_{x\to s_i} \,\, K(x_0,x) K(x_1,x) {\partial \over \partial x}\Big(      \cr
&& 2W_{1+n}^{(g-1)}(x,x,J)  + 2 \sum_{h}\sum'_{I\in J}\,\, W_{n-|I|}^{(g-h)}(x,J/I)     W_{1+|I|}^{(h)}(x,I)  
\,\, \Big) \cr
\eea
The last term can be integrated by parts, and we get:
\bea
&& W_{n+1}^{(g)}(x_0,x_1,J)-W_{n+1}^{(g)}(x_1,x_0,J) \cr
&=& 2 \sum_i \Res_{x\to s_i} \left( K(x_0,x)\,\,  B(x,x_1)  - K(x_1,x)\,\,  B(x,x_0) \right) \, W_{n}^{(g)}(x,J) \cr
&& +{1\over 2}  \sum_{i} \Res_{x\to s_i} \,\, \Big( K'(x_0,x) K(x_1,x)-K(x_0,x) K'(x_1,x)\Big)\,\, 
\Big(     \cr
&& 2W_{1+n}^{(g-1)}(x,x,J)  + 2 \sum_{h}\sum'_{I\in J}\,\, W_{n-|I|}^{(g-h)}(x,J/I)     W_{1+|I|}^{(h)}(x,I)  
\Big)\,\,  \cr
\eea
Then we use theorem \ref{thWngPng}:
\bea
&& W_{n+1}^{(g)}(x_0,x_1,J)-W_{n+1}^{(g)}(x_1,x_0,J) \cr
&=& 2 \sum_i \Res_{x\to s_i} \left( K(x_0,x)\,\,  B(x,x_1)  - K(x_1,x)\,\,  B(x,x_0) \right) \, W_{n}^{(g)}(x,J) \cr
&& + \sum_{i} \Res_{x\to s_i} \,\, \Big( K'(x_0,x) K(x_1,x)-K(x_0,x) K'(x_1,x)\Big)\,\, 
\Big(      P_{n}^{(g)}(x,J) \cr
&& + (Y(x) - \hbar \partial_x) W_{n}^{(g)}(x,J)  + \sum_j \partial_{x_j}
\Big( {W_{n-1}^{(g)}(x_j,J/\{x_j\})\over x-x_j} \Big)
\Big)\,\,  \cr
\eea
Since $P_{n}^{(g)}(x,J)$ and $W_{n-1}^{(g)}(x_j,J/\{x_j\})$ have no poles at the $s_i$'s, we have:
\bea
&& W_{n+1}^{(g)}(x_0,x_1,J)-W_{n+1}^{(g)}(x_1,x_0,J) \cr
&=& 2 \sum_i \Res_{x\to s_i} \left( K(x_0,x)\,\,  B(x,x_1)  - K(x_1,x)\,\,  B(x,x_0) \right) \, W_{n}^{(g)}(x,J) \cr
&+& \sum_{i} \Res_{x\to s_i} \,\, \Big( K'(x_0,x) K(x_1,x)-K(x_0,x) K'(x_1,x)\Big)\cr
&&(Y(x) - \hbar \partial_x) W_{n}^{(g)}(x,J)    \cr
\eea
Notice that:
\beq
K'_0 K_1 - K_0 K'_1 = -{1\over \hbar}\,( G_0 K_1 - K_0 G_1)
\eeq
and $B=-{1\over 2}\, G'$, therefore:
\bea
&& W_{n+1}^{(g)}(x_0,x_1,J)-W_{n+1}^{(g)}(x_1,x_0,J) \cr
&=& - \sum_i \Res_{x\to s_i} \left( K_0 G'_1  - K_1 G'_0 \right) \, W_{n}^{(g)}(x,J) \cr
&& -{1\over \hbar}  \sum_{i} \Res_{x\to s_i} \,\, \Big( G_0 K_1- K_0 G_1 \Big)\,\, 
(Y(x) - \hbar \partial_x) W_{n}^{(g)}(x,J)    \cr
\eea
we integrate the first line by parts:
\bea
&& W_{n+1}^{(g)}(x_0,x_1,J)-W_{n+1}^{(g)}(x_1,x_0,J) \cr
&=& \sum_i \Res_{x\to s_i} \left( K'_0 G_1  - K'_1 G_0 \right) \, W_{n}^{(g)}(x,J) \cr
&& + \sum_i \Res_{x\to s_i} \left( K_0 G_1  - K_1 G_0 \right) \, W_{n}^{(g)}(x,J)' \cr
&& -{1\over \hbar}  \sum_{i} \Res_{x\to s_i} \,\, \Big( G_0 K_1- K_0 G_1 \Big)\,\, 
(Y(x) - \hbar \partial_x) W_{n}^{(g)}(x,J)    \cr
\eea
Notice that:
\beq
K'_0 G_1 - G_0 K'_1 = -{Y\over \hbar}\,( K_0 G_1 - G_0 K_1)
\eeq
So we find \beq  W_{n+1}^{(g)}(x_0,x_1,J)-W_{n+1}^{(g)}(x_1,x_0,J)=0\eeq

\appendix{Proof of theorem \ref{thWngindeptK}}
\label{approofthWngindeptK}

\bt
The correlation functions $W_n^{(g)}$ are independent of the choice of kernel $K$, provided that $K$ is solution of the equation \eq{diffeqdefK}.
\et

\proof{

Any two solutions of  \eq{diffeqdefK}, differ by a homogeneous solution, i.e. by $\psi^2(x)$.
Therefore, what we have to prove is that the following quantity vanishes:
\beq
\sum_i \Res_{x\to s_i} \psi^2(x)\, \Big[ W_{n+2}^{(g-1)}(x,x,J) + \sum_h \sum'_{I\subset J}\, W_{1+|I|}^{(h)}(x,I) W_{1+n-|I|}^{(g-h)}(x,J/I) \Big]
\eeq
Using theorem \ref{thWngPng}, we have:
\bea
&& \Res_{x\to s_i} \psi^2(x)\, \Big[ W_{n+2}^{(g-1)}(x,x,J) + \sum_h \sum'_{I\subset J}\, W_{1+|I|}^{(h)}(x,I) W_{1+n-|I|}^{(g-h)}(x,J/I) \Big] \cr
&=& \Res_{x\to s_i} \psi^2(x) \Big(Y(x)W_n^{(g)}(x,J) - \hbar \partial_x W_n^{(g)}(x,J)+P_n^{(g)}(x;J) \Big) 
\eea
Then we notice that $P_n^{(g)}$ gives no residue, and then we use $Y=-2\hbar \psi'/\psi$, and we integrate by parts:
\bea
&=& -\hbar \Res_{x\to s_i} \psi^2(x) \Big(2\,{\psi'\over \psi}\,W_n^{(g)} + \partial_x W_n^{(g)} \Big) \cr
&=& -\hbar \Res_{x\to s_i} \partial_x\, \Big(\psi^2\,\,  W_n^{(g)} \Big) \cr
&=& 0
\eea
This means that adding to $K(x_0,x)$ a constant times $\psi^2(x)$ doesnot change the $W_n^{(g)}$'s.
In fact we may chose a different constant near each $s_i$, or in other words, we may assume that
\beq\label{Ki2vanishhyp}
K_{i,2}(x_0)=0
\eeq

}


\appendix{Proof of theorem \ref{thW3Krich}}
\label{approofthW3Krich}

{\bf Theorem \ref{thpolessiWng}}
{\it 
The 3 point function $W_3^{(0)}$ is symmetric and we have:
\beq
W_3^{(0)}(x_1,x_2,x_3) = 4\,\sum_i \Res_{x\to s_i}\,\, {B(x,x_1)B(x,x_2)B(x,x_3)\over Y'(x)}
\eeq
}
\bigskip

{\bf proof:}

The definition of $W_3^{(0)}$ is:
\bea  
&& W_3^{(0)}(x_0,x_1,x_2)\cr 
&=& 2 \sum_i \Res_{x\to s_i} K(x_0,x)B(x,x_1)B(x,x_2) \cr
 &=& {1 \over 2} \sum_i \Res_{x\to s_i}  K_0 \, G_1^{'} \, G'_2 \cr
 &=& {1 \over 2} \sum_i \Res_{x\to s_i} K_0 \left( (\hbar K''_1 +  Y K'_1 +  Y'
 K_1)(\hbar K''_2 + YK'_2 +Y' K_2) \right) \cr
 &=& {1 \over 2} \sum_i \Res_{x\to s_i} K_0 \, (\, \hbar^2 K''_1 K''_2 +  \hbar Y (K'_1
 K''_2+K''_1 K'_2) +  \hbar Y' (K''_1 K_2 +K''_2 K_1) \cr
 && +  Y^2 K'_1 K'_2+  Y Y' (K_1 K'_2+K'_1 K_2)+
{Y'}^2 K_1 K_2 \,) \cr
 \eea
where we have written for short $K_i = K(x_i,x)$, $G_i=G(x_i,x)$, and derivative are w.r.t. $x$.

Since $K(x_i,x)$ has no pole when $x\to s_i$,
the first term vanishes.
Using the Ricatti equation $Y^2 = 2 \hbar Y' + U$ (where $U$ has no pole at $s_i$), we may replace $Y^2$ by $2 \hbar Y'$ and $ Y Y'$ by $\hbar Y''$ without changing the residues, i.e.:
\bea
& & W_3^{(0)}(x_0,x_1,x_2)\cr
 &=& {1 \over 2} \sum_i \Res_{x\to s_i} K_0 \, (  \hbar Y (K'_1
 K''_2+K''_1 K'_2) +  \hbar Y' (K''_1 K_2 +K''_2 K_1) \cr
 && + 2 \hbar Y' K'_1 K'_2+ \hbar Y'' (K_1 K'_2+K'_1 K_2)+
{Y'}^2 K_1 K_2 \,)\ \cr
 &=&  {1\over 2} \sum_i \Res_{x\to s_i} K_0 \, (  \hbar Y (K'_1 K'_2)' +  \hbar Y' (K_1 K_2)'' + \hbar Y'' (K_1 K_2)'+  {Y'}^2 K_1 K_2 \,)\ \cr
 &=&  {1\over 2} \sum_i \Res_{x\to s_i} {Y'}^2 K_0 K_1 K_2 +
 \hbar \big( Y'' K_0 (K_1 K_2)' - (Y K_0)' K'_1 K'_2 - (Y' K_0)' (K_1 K_2)' \big) \cr
 &=&  {1\over 2} \sum_i \Res_{x\to s_i} {Y'}^2 K_0 K_1 K_2 -
 \hbar \big(  (Y K_0)' K'_1 K'_2 + Y' K_0' (K_1 K_2)' \big) \cr
&=&  {1\over 2} \sum_i \Res_{x\to s_i} {Y'}^2 K_0 K_1 K_2 - \hbar Y K'_0 K'_1 K'_2 - \hbar Y' (K_0 K'_1 K'_2+K'_0 K_1 K'_2+K'_0 K'_1 K_2) \cr
\eea
This expression is clearly symmetric in $x_0, x_1, x_2$ as claimed in theorem \ref{thsym}.

Let us give an alternative expression, in the form of the Verlinde or Krichever formula \cite{Krich}:
\beq\label{eq30Krichever}
W_3^{(0)}(x_1,x_2,x_3)=
4 \sum_i \Res_{x\to s_i} \, {B(x,x_1)B(x,x_2)B(x,x_3) \over Y^{'}(x)}
\eeq

{\bf proof:}

In order to prove formula \ref{eq30Krichever}, compute:
\beq
B(x,x_i) = -{1\over 2} G'(x,x_i) = -{1\over 2} G'_i = {1\over 2}(\hbar K''_i +  Y K'_i +  Y' K_i)
\eeq
thus:
\bea
&&  \sum_i \Res_{x\to s_i} \, {B(x,x_1)B(x,x_2)B(x,x_3) \over Y^{'}(x)} \cr
&=&  {1\over 8}\sum_i \Res_{x\to s_i} \, {1\over Y'(x)}\, (\hbar K''_0 +  Y K'_0 +  Y' K_0)(\hbar K''_1 +  Y K'_1 +  Y' K_1)\cr
&& \qquad \qquad (\hbar K''_2 +  Y K'_2 +  Y' K_2)  \cr
&=&  {1\over 8}\sum_i \Res_{x\to s_i} \,
 {\hbar^3\over Y'} K''_0 K''_1 K''_2 + \hbar^2 {Y\over Y'} (K'_0 K''_1 K''_2 + K''_0 K'_1 K''_2 + K''_0 K''_1 K'_2) \cr
&& + \hbar^2  (K_0 K''_1 K''_2 + K''_0 K_1 K''_2 + K''_0 K''_1 K_2) \cr
&& + \hbar {Y^2\over Y'} (K''_0 K'_1 K'_2+K'_0 K''_1 K'_2+K'_0 K'_1 K''_2) \cr
&& + \hbar Y (K_0 K'_1 K''_2 + K_0 K''_1 K'_2 + K'_0 K_1 K''_2 + K'_0 K''_1 K_2 + K''_0 K_1 K'_2 + K''_0 K'_1 K_2) \cr
&& + \hbar Y' (K''_0 K_1 K_2 + K_0 K''_1 K_2 + K_0 K_1 K''_2)
+ {Y^3\over Y'} K'_0 K'_1 K'_2  \cr
&& + Y^2  (K_0 K'_1 K'_2 + K'_0 K_1 K'_2 + K'_0 K'_1 K_2) \cr
&& + Y Y' (K'_0 K_1 K_2 + K_0 K'_1 K_2 + K_0 K_1 K'_2)  + Y'^2 K_0 K_1 K_2  \cr
\eea
Notice that $K_i$ has no pole at the $s_i$'s, and $1/Y'$ has no pole, $Y/Y'$ has no pole, $Y^2/Y'$ has no pole, thus:
\bea
&&  \sum_i \Res_{x\to s_i} \, {B(x,x_1)B(x,x_2)B(x,x_3) \over Y^{'}(x)} \cr
&=&  {1\over 8}\sum_i \Res_{x\to s_i} \,
 \hbar Y (K_0 K'_1 K''_2 + K_0 K''_1 K'_2 + K'_0 K_1 K''_2 + K'_0 K''_1 K_2 + K''_0 K_1 K'_2 \cr
 && + K''_0 K'_1 K_2) 
 + \hbar Y' (K''_0 K_1 K_2 + K_0 K''_1 K_2 + K_0 K_1 K''_2)
+ {Y^3\over Y'} K'_0 K'_1 K'_2  \cr
&& + Y^2  (K_0 K'_1 K'_2 + K'_0 K_1 K'_2 + K'_0 K'_1 K_2) \cr
&& + Y Y' (K'_0 K_1 K_2 + K_0 K'_1 K_2 + K_0 K_1 K'_2)  + Y'^2 K_0 K_1 K_2  \cr
\eea
Notice that $Y^2 = 2\hbar Y' + U$, thus we may replace $Y^3/Y'$ by $2\hbar Y$, and $Y^2$ by $2\hbar Y'$ and $Y Y'$ by $\hbar Y''$, thus:
\bea
&&  \sum_i \Res_{x\to s_i} \, {B(x,x_1)B(x,x_2)B(x,x_3) \over Y^{'}(x)} \cr
&=&  {1\over 8}\sum_i \Res_{x\to s_i} \,
 \hbar Y (K_0 K'_1 K''_2 + K_0 K''_1 K'_2 + K'_0 K_1 K''_2 + K'_0 K''_1 K_2 + K''_0 K_1 K'_2 \cr
 && + K''_0 K'_1 K_2) 
 + \hbar Y' (K''_0 K_1 K_2 + K_0 K''_1 K_2 + K_0 K_1 K''_2)
+ 2\hbar Y K'_0 K'_1 K'_2  \cr
&& + 2 \hbar Y'  (K_0 K'_1 K'_2 + K'_0 K_1 K'_2 + K'_0 K'_1 K_2)
+ \hbar Y'' (K'_0 K_1 K_2 + K_0 K'_1 K_2 + K_0 K_1 K'_2) \cr
&& + Y'^2 K_0 K_1 K_2  \cr
&=&  {1\over 8}\sum_i \Res_{x\to s_i} \,
 \hbar Y (K_0 (K'_1 K'_2)' +  K_1 (K'_0 K'_2)' + K_2 (K'_0 K'_1)') \cr
&& + 2\hbar Y K'_0 K'_1 K'_2  + Y'^2 K_0 K_1 K_2  + \hbar (Y' (K'_0 K_1 K_2 + K_0 K'_1 K_2 + K_0 K_1 K'_2))' \cr
&=&  {1\over 8}\sum_i \Res_{x\to s_i} \,
 \hbar Y (K_0 (K'_1 K'_2)' +  K_1 (K'_0 K'_2)' + K_2 (K'_0 K'_1)') \cr
&& + 2\hbar Y K'_0 K'_1 K'_2  + Y'^2 K_0 K_1 K_2  \cr
&=&  -{1\over 8}\sum_i \Res_{x\to s_i} \, 3 \hbar Y K'_0 K'_1 K'_2 + \hbar Y' (K_0 K'_1 K'_2 + K'_0 K_1 K'_2 + K'_0 K'_1 K_2) \cr
&& - 2\hbar Y K'_0 K'_1 K'_2  - Y'^2 K_0 K_1 K_2  \cr
&=& {1\over 4} W_3^{(0)}(x_0,x_1,x_2)
\eea

\subsection{Direct computation}

We write 
\bea
&& W_3^{(0)}(z_1,z_2,z_3) \cr
&=& 2 \sum_i \Res_{z\to s_i} K(z_1,z)B(z_2,z)B(z_3,z) \cr
&=& \sum_j\sum_i {A_{i,j}\over (z_2-s_j)^2}\, \Res_{z\to s_i} K(z_1,z){1\over (z-s_i)^2(z_3-z)^2} \,\,+{\rm sym.} \cr
&& + 2 \sum_i\sum_{i'\neq i} \sum_{j,k} {A_{i,j}A_{i',k}\over (z_2-s_j)^2(z_3-s_k)^2}\, \Res_{z\to s_i} K(z_1,z) {1\over (z-s_i)^2(z-s_{i'})^2}  \,\,+{\rm sym.}  \cr
&& + 2 \sum_i \sum_{j,k} {A_{i,j}A_{i,k}\over (z_2-s_j)^2(z_3-s_k)^2}\, \Res_{z\to s_i} K(z_1,z) {1\over (z-s_i)^4}    \cr
&=& \sum_j\sum_i {A_{i,j}\over (z_2-s_j)^2}\, \Big( {K_{i,1}(z_1)\over (z_3-s_i)^2}+{2 K_{i,0}(z_1)\over (z_3-s_i)^3}  \Big)\,\,+{\rm sym.} \cr
&& + 2 \sum_i\sum_{i'\neq i} \sum_{j,k} {A_{i,j}A_{i',k}\over (z_2-s_j)^2(z_3-s_k)^2}\, \Big( {K_{i,1}(z_1)\over (s_{i'}-s_i)^2}+{2 K_{i,0}(z_1)\over (s_{i'}-s_i)^3}  \Big) \,\,+{\rm sym.}  \cr
&& + 2 \sum_i \sum_{j,k} {A_{i,j}A_{i,k}\over (z_2-s_j)^2(z_3-s_k)^2}\, K_{i,3}(z_1)   \cr
&=& \sum_j\sum_i {A_{i,j}\over (z_2-s_j)^2}\, \Big( {K_{i,1}(z_1)\over (z_3-s_i)^2}+{2 K_{i,0}(z_1)\over (z_3-s_i)^3}  \Big)\,\,+{\rm sym.} \cr
&& + 2 \sum_i\sum_{i'\neq i} \sum_{j,k} {A_{i,j}A_{i',k}\over (z_2-s_j)^2(z_3-s_k)^2}\, \Big( {K_{i,1}(z_1)\over (s_{i'}-s_i)^2}+{2 K_{i,0}(z_1)\over (s_{i'}-s_i)^3}  \Big) \,\,+{\rm sym.}  \cr
&& - 2 \sum_i \sum_{j,k} {A_{i,j}A_{i,k}\over (z_2-s_j)^2(z_3-s_k)^2}\, T_{i,i}\, K_{i,1}(z_1)   \cr
&& - 2 \sum_i \sum_{j,k} {A_{i,j}A_{i,k}\over (z_2-s_j)^2(z_3-s_k)^2}\, ({V'''(s_i)\over 2\hbar}+2\sum_{i'\neq i} {1\over (s_{i'}-s_i)^3}) K_{i,0}(z_1)   \cr
&& + {2\over \hbar} \sum_i \sum_{j,k} {A_{i,j}A_{i,k}\over (z_2-s_j)^2(z_3-s_k)^2(z_1-s_i)^3}   \cr
&& + {4\over \hbar} \sum_i\sum_{i'\neq i}\sum_l \sum_{j,k} {A_{i,j}A_{i,k} A_{i',l}\over (z_2-s_j)^2(z_3-s_k)^2(s_{i'}-s_i)^3(z_1-s_l)^2}\,   \cr
&=& {2\over \hbar} \sum_{i,j,k} {A_{i,j}A_{i,k} \over (z_1-s_i)^3(z_2-s_j)^2(z_3-s_k)^2} + {A_{j,i}A_{j,k} \over (z_1-s_i)^2(z_2-s_j)^3(z_3-s_k)^2} \cr
&&+ {A_{k,i}A_{k,j} \over (z_1-s_i)^2(z_2-s_j)^2(z_3-s_k)^3}
 \cr
&& + \sum_{i,j,k} {K_{i,1}(z_1)\over (z_2-s_j)^2(z_3-s_k)^2}\,\Big( A_{j,k}\delta_{i,j}+A_{j,k}\delta_{i,k}  - A_{i,j} \sum_{i'} T_{i,i'} A_{i',k} \cr
&& - A_{i,k} \sum_{i'} T_{i,i'} A_{i',j}  \Big) \cr
&& + 2 \sum_i\sum_{i'\neq i} \sum_{j,k} {A_{i,j}A_{i',k}\over (z_2-s_j)^2(z_3-s_k)^2}\, {2 K_{i,0}(z_1)\over (s_{i'}-s_i)^3}   \,\,+{\rm sym.}  \cr
&& - 2 \sum_i \sum_{j,k} {A_{i,j}A_{i,k}\over (z_2-s_j)^2(z_3-s_k)^2}\, ({V'''(s_i)\over 2\hbar}+2\sum_{i'\neq i} {1\over (s_{i'}-s_i)^3}) K_{i,0}(z_1)   \cr
&& + {4\over \hbar} \sum_i\sum_{i'\neq i}\sum_l \sum_{j,k} {A_{i,j}A_{i,k} A_{i',l}\over (z_2-s_j)^2(z_3-s_k)^2(s_{i'}-s_i)^3(z_1-s_l)^2}\,   \cr
&=& {2\over \hbar} \sum_{l,j,k} {1 \over (z_1-s_l)^2(z_2-s_j)^2(z_3-s_k)^2} \sum_i \Big( {\delta_{i,l} A_{i,j} A_{i,k}\over (z_1-s_i)}+{\delta_{i,j} A_{i,l} A_{i,k}\over (z_2-s_i)}\cr
&& +{\delta_{i,k} A_{i,l} A_{i,j}\over (z_3-s_i)}  \Big) \cr
 && + {4\over \hbar}\sum_{l,j,k}\sum_i \sum_{i'\neq i} {A_{i,j}A_{i,k} A_{i',l} +  A_{i,j}A_{i',k} A_{i,l}  +  A_{i,k}A_{i',j} A_{i,l} -  A_{i,j}A_{i,k}A_{i,l}\over (z_1-s_l)^2(z_2-s_j)^2(z_3-s_k)^2(s_{i'}-s_i)^3}\,\cr
 && - {1\over \hbar^2}\sum_{l,j,k}\sum_i  {A_{i,j}A_{i,k}A_{i,l} V'''(s_i)\over (z_1-s_l)^2(z_2-s_j)^2(z_3-s_k)^2}\,\cr%
\eea

Thus we have:
\bea\label{W30exact}
&& W_3^{(0)}(z_1,z_2,z_3) \cr
&=& {2\over \hbar} \sum_{i,j,k,l} { {\delta_{i,l} A_{i,j} A_{i,k}\over (z_1-s_i)}+{\delta_{i,j} A_{i,l} A_{i,k}\over (z_2-s_i)}+{\delta_{i,k} A_{i,l} A_{i,j}\over (z_3-s_i)}  \over (z_1-s_l)^2(z_2-s_j)^2(z_3-s_k)^2}  \cr
 && + {4\over \hbar}\sum_{l,j,k}\sum_i \sum_{i'\neq i} {A_{i,j}A_{i,k} A_{i',l} +  A_{i,j}A_{i',k} A_{i,l}  +  A_{i,k}A_{i',j} A_{i,l} -  A_{i,j}A_{i,k}A_{i,l}\over (z_1-s_l)^2(z_2-s_j)^2(z_3-s_k)^2(s_{i'}-s_i)^3}\,\cr
 && - {1\over \hbar^2}\sum_{l,j,k}\sum_i  {A_{i,j}A_{i,k}A_{i,l} V'''(s_i)\over (z_1-s_l)^2(z_2-s_j)^2(z_3-s_k)^2}\,\cr%
\eea

\appendix{Proof of theorem \ref{thvariationV}}
\label{approofthvariationV}

{\bf Theorem \ref{thvariationV}}
{\it 
Under an infinitesimal variation of the potential $V\to V+\delta V$, we have:
\beq
\forall n\geq 0, g\geq 0
\,\, , \quad
\delta W_{n}^{(g)}(x_1,\dots,x_n) = - \sum_i \Res_{x\to s_i} W_{n+1}^{(g)}(x,x_1,\dots,x_n)\, \delta V(x)
\eeq
}
\bigskip

\subsection{Variation of $\om$}

We have:
\beq
\om(x) = \hbar \sum_i {1\over x-s_i}
\eeq
and
\beq
V'(s_i) = 2\hbar \sum_{j\neq i} {1\over s_i-s_j}
\eeq
Thus taking a variation we have:
\beq
\delta V'(s_i) + \delta s_i V''(s_i)  = -2\hbar \sum_{j\neq i} {\delta s_i-\delta s_j\over (s_i-s_j)^2}
\eeq
i.e.
\beq
\delta V'(s_i) = -\hbar \sum_j T_{i,j} \delta s_j
\eeq
which implies:
\beq
\delta s_i = - {1\over \hbar}\sum_j A_{i,j} \delta V'(s_j)
\eeq
and therefore:
\beq
\delta \om(x) = -\sum_{i,j} {A_{i,j} \delta V'(s_j)\over (x-s_i)^2}
\eeq
which can also be written:
\bea
\delta \om(x) 
&=& - \sum_k \Res_{x'\to s_k} \sum_{i,j} \,\, {A_{i,j} \over (x-s_i)^2\, (x'-s_j)}\,\, \delta V'(x') \cr
&=& - \sum_k \Res_{x'\to s_k} \sum_{i,j} \,\, {A_{i,j} \over (x-s_i)^2\, (x'-s_j)^2}\,\, \delta V(x') \cr
&=& - \sum_k \Res_{x'\to s_k} \, B(x,x')\,\, \delta V(x') 
\eea
and finally we obtain the case $n=1,g=0$ of the theorem:
\beq\label{varomResB}
\encadremath{
\delta \om(x)  = - \sum_k \Res_{x'\to s_k} \, B(x,x')\,\, \delta V(x') 
}\eeq

\subsection{Variation of $B$}

Consider:
\beq
\ovl{W}_2^{(0)}(x,x') = B(x,x') - {1\over 2}\,{1\over (x-x')^2} =  \sum_{i,j} {A_{i,j}\over (x-s_j)^2(x'-s_i)^2} 
\eeq
Due to \eq{eqmono1} we have:
\bea
\ovl{W}_2^{(0)}(x,x')
&=&  \sum_{i} {\hbar K(x,s_i)\over (x'-s_i)^2} \cr
&=&  \sum_i \Res_{z\to s_i} K(x,z)\,\, {\om(z)\over (z-x')^2} \cr
&=&  {\partial \over \partial x'}\, \sum_i \Res_{z\to s_i} K(x,z)\,\, {\om(z)-\om(x')\over z-x'} \cr
\eea
On the other hand, since $\ovl{W}_2^{(0)}(x,x')$ has poles only at the $s_i$'s we have:
\bea
\ovl{W}_2^{(0)}(x,x')
&=& \Res_{z\to x} G(x,z)\,\, \ovl{W}_2^{(0)}(z,x') \cr
&=& - \sum_i \Res_{z\to s_i} G(x,z)\,\, \ovl{W}_2^{(0)}(z,x') \cr
&=& - \sum_i \Res_{z\to s_i} \left( (2\om(z)-V'(z)+\hbar {\partial_z})K(x,z)\right) \,\, \ovl{W}_2^{(0)}(z,x') \cr
&=& - \sum_i \Res_{z\to s_i} K(x,z)\,\, \left( (2\om(z)-V'(z)-\hbar {\partial_z}) \,\, \ovl{W}_2^{(0)}(z,x') \right)\cr
\eea
This implies that $\forall x$:
\beq
0= - \sum_i \Res_{z\to s_i} K(x,z)\,\, \left( (2\om(z)-V'(z)-\hbar {\partial_z}) \,\, \ovl{W}_2^{(0)}(z,x') + {\partial \over \partial x'} {\om(z)-\om(x')\over z-x'} \right) 
\eeq
and therefore, $\ovl{W}_2^{(0)}(x,x')$ satisfies the loop equation:
\beq\label{apploopeqW20}
(2\om(x)-V'(x)-\hbar {\partial_x}) \,\, \ovl{W}_2^{(0)}(x,x') + {\partial \over \partial x'} {\om(x)-\om(x')\over x-x'} = - P_2^{(0)}(x,x')
\eeq
where $P_2^{(0)}(x,x')$ has no pole at  $x\to s_i$'s.

\medskip
Then we take the variation:
\bea
(2\om(x)-V'(x)-\hbar {\partial_x}) \,\, \delta \ovl{W}_2^{(0)}(x,x') 
&=&
- (2\delta \om(x)-\delta V'(x)) \,\, \ovl{W}_2^{(0)}(x,x')\cr
&& - {\partial \over \partial x'} {\delta \om(x)-\delta \om(x')\over x-x'}  - \delta P_2^{(0)}(x,x') \cr
\eea
$\delta \ovl{W}_2^{(0)}(x,x')$ is a rational fraction of $x$, with poles only at the $s_i$'s, and $\delta P_2^{(0)}(x,x')$ has no pole at  $x\to s_i$'s.
We thus write:
\bea
\delta W_2^{(0)}(x,x')&=&\delta \ovl{W}_2^{(0)}(x,x')\cr
&=& \Res_{z\to x} G(x,z)\,\, \delta \ovl{W}_2^{(0)}(z,x') \cr
&=& - \sum_i \Res_{z\to s_i} G(x,z)\,\, \ovl{W}_2^{(0)}(z,x') \cr
&=& - \sum_i \Res_{z\to s_i} \left( (2\om(z)-V'(z)+\hbar {\partial_z})K(x,z)\right) \,\, \delta \ovl{W}_2^{(0)}(z,x') \cr
&=& - \sum_i \Res_{z\to s_i} K(x,z)\,\, \left( (2\om(z)-V'(z)-\hbar {\partial_z}) \,\, \delta \ovl{W}_2^{(0)}(z,x') \right)\cr
&=&  \sum_i \Res_{z\to s_i} K(x,z)\,\, \Big(  (2\delta \om(z)-\delta V'(z)) \,\, \ovl{W}_2^{(0)}(z,x') \cr
&& + {\partial \over \partial x'} {\delta \om(z)-\delta \om(x')\over z-x'}  + \delta P_2^{(0)}(z,x') \Big)\cr
&=&  \sum_i \Res_{z\to s_i} K(x,z)\,\, \Big(  (2\delta \om(z)-\delta V'(z)) \,\, \ovl{W}_2^{(0)}(z,x') 
 + {\delta \om(z)\over (z-x')^2}   \Big)\cr
&=&  \sum_i \Res_{z\to s_i} K(x,z)\,\,   (2\delta \om(z)-\delta V'(z)) \,\, B(z,x')   \cr
\eea
Then, we use \eq{varomResB}, and we get:
\bea
\delta W_2^{(0)}(x,x')
&=& - 2  \sum_i \Res_{z\to s_i} \sum_k \Res_{x''\to s_k} K(x,z)\,\,  B(z,x'') \delta V(x'')   \,\, B(z,x')   \cr
&& -  \sum_i \Res_{z\to s_i} K(x,z)\,\,   \delta V'(z) \,\, B(z,x')   \cr
&=& -   \sum_i \Res_{z\to s_i} \sum_k \Res_{x''\to s_k} K(x,z)\,\,  G(z,x'') \delta V'(x'')   \,\, B(z,x')   \cr
&& -   \sum_i \Res_{z\to s_i} \Res_{x''\to z} K(x,z)\,\,  G(z,x'') \delta V'(x'')   \,\, B(z,x')   \cr
&=& -   \sum_k \Res_{x''\to s_k} \sum_i \Res_{z\to s_i} K(x,z)\,\,  G(z,x'') \delta V'(x'')   \,\, B(z,x')   \cr
&=& - 2   \sum_k \Res_{x''\to s_k} \sum_i \Res_{z\to s_i} K(x,z)\,\,  B(z,x'') \delta V(x'')   \,\, B(z,x')   \cr
\eea
We thus obtain the case $n=2,g=0$ of the theorem:
\beq\label{varBResW3}
\encadremath{
\delta W_2^{(0)}(x,x') =  - \sum_k \Res_{x''\to s_k} W_3^{(0)}(x,x',x'')\,\, \delta V(x'')    
}\eeq

\subsection{Variation of other higher correlators}

We prove by recursion on $2g+n$, that:
\beq
\encadremath{
\delta W^{(g)}_{n+1}(x,L) 
= - \sum_k \Res_{x''\to s_k}\,  \delta V(x'')\,\,  W_{n+2}^{(g)}(z,L,x'')  
}\eeq
where $L=\{x_1,\dots,x_n\}$.

\medskip

We write:
\beq\label{defUng2}
U_{n+1}^{(g)}(z,L) = \ovl{W}_{n+2}^{(g-1)}(z,z,L) + \sum_h\sum'_{J\subset L} W^{(h)}_{1+|J|}(z,J)\, W^{(g-h)}_{1+n-|J|}(z,L/J)
\eeq
By definition we have:
\beq
W^{(g)}_{n+1}(x,L) = \sum_i \Res_{z\to s_i} K(x,z)\,\, U^{(g)}_{n+1}(z,L)
\eeq

From the recursion hypothesis, we have:
\bea
\delta U_{n+1}^{(g)}(z,L) 
& = & - \sum_k \Res_{x''\to s_k}\, \delta V(x'')\,\, \Big( W_{n+3}^{(g-1)}(z,z,L,x'') \cr
&& - 2 \sum_h\sum'_{J\subset L}{W}^{(h)}_{2+|J|}(z,J,x'')\, {W}^{(g-h)}_{1+n-|J|}(z,L/J) \Big) \cr
& = & - \sum_k \Res_{x''\to s_k}\, \delta V(x'')\,\, \Big( U_{n+2}^{(g)}(z,L,x'') - 2 B(z,x'') W_{n+1}^{(g)}(z,L)
 \Big) \cr
\eea

Thus:
\bea
&& \delta W^{(g)}_{n+1}(x,L) \cr
&=& \sum_i \Res_{z\to s_i} \delta K(x,z)\,\, U^{(g)}_{n+1}(z,L)  - \sum_i \Res_{z\to s_i} K(x,z)\,\, \sum_k \Res_{x''\to s_k}\, \delta V(x'')\,\, \Big( \cr
&&  U_{n+2}^{(g)}(z,L,x'')  - 2 B(z,x'') W_{n+1}^{(g)}(z,L) \Big) \cr
&=& \sum_i \Res_{z\to s_i} \delta K(x,z)\,\, U^{(g)}_{n+1}(z,L)  - \sum_k \Res_{x''\to s_k}\, \sum_i \Res_{z\to s_i} K(x,z)\,\,  \delta V(x'')\,\, \Big( \cr
&& U_{n+2}^{(g)}(z,L,x'') - 2 B(z,x'') W_{n+1}^{(g)}(z,L) \Big) \cr
&=& \sum_i \Res_{z\to s_i} \delta K(x,z)\,\, U^{(g)}_{n+1}(z,L) \cr
&& +2 \sum_k \Res_{x''\to s_k}\, \sum_i \Res_{z\to s_i} K(x,z)\,\,  \delta V(x'')\,\,  B(z,x'') W_{n+1}^{(g)}(z,L)  \cr
&& - \sum_k \Res_{x''\to s_k}\, \sum_i \Res_{z\to s_i} K(x,z)\,\,  \delta V(x'')\,\,  U_{n+2}^{(g)}(z,L,x'')  \cr
&=& \sum_i \Res_{z\to s_i} \delta K(x,z)\,\, U^{(g)}_{n+1}(z,L) \cr
&& +2 \sum_i \Res_{z\to s_i}\sum_k \Res_{x''\to s_k}\,  K(x,z)\,\,  \delta V(x'')\,\,  B(z,x'') W_{n+1}^{(g)}(z,L)  \cr
&& +2 \, \sum_i \Res_{z\to s_i} \Res_{x''\to z} K(x,z)\,\,  \delta V(x'')\,\,  B(z,x'') W_{n+1}^{(g)}(z,L)  \cr
&& - \sum_k \Res_{x''\to s_k}\,  \delta V(x'')\,\,  W_{n+2}^{(g)}(z,L,x'')  \cr
\eea
We use the loop equation of theorem \ref{thWngPng}, which says that
$U^{(g)}_{n+1}(z,L) + (2\om(z)-V'(z)+\hbar \partial_z)W^{(g)}_{n+1}(z,L) $ has no pole at $z\to s_i$, and thus:
\bea
&& \delta W^{(g)}_{n+1}(x,L) \cr
&=& - \sum_i \Res_{z\to s_i} \delta K(x,z)\,\, (2\om(z)-V'(z)+\hbar \partial_z)W^{(g)}_{n+1}(z,L) \cr
&& +2 \sum_i \Res_{z\to s_i}\sum_k \Res_{x''\to s_k}\,  K(x,z)\,\,  \delta V(x'')\,\,  B(z,x'') W_{n+1}^{(g)}(z,L)  \cr
&& +2 \, \sum_i \Res_{z\to s_i} \Res_{x''\to z} K(x,z)\,\,  \delta V(x'')\,\,  B(z,x'') W_{n+1}^{(g)}(z,L)  \cr
&& - \sum_k \Res_{x''\to s_k}\,  \delta V(x'')\,\,  W_{n+2}^{(g)}(z,L,x'')  \cr
&=& - \sum_i \Res_{z\to s_i} W^{(g)}_{n+1}(z,L) \,\, (2\om(z)-V'(z)-\hbar \partial_z) \delta K(x,z) \cr
&& +2 \sum_i \Res_{z\to s_i}\sum_k \Res_{x''\to s_k}\,  K(x,z)\,\,  \delta V(x'')\,\,  B(z,x'') W_{n+1}^{(g)}(z,L)  \cr
&& +2 \, \sum_i \Res_{z\to s_i} \Res_{x''\to z} K(x,z)\,\,  \delta V(x'')\,\,  B(z,x'') W_{n+1}^{(g)}(z,L)  \cr
&& - \sum_k \Res_{x''\to s_k}\,  \delta V(x'')\,\,  W_{n+2}^{(g)}(z,L,x'')  \cr
\eea
and we have:
\beq
(2\om(z)-V'(z)-\hbar \partial_z) \delta K(x,z)
 = \delta G(x,z) - (2\delta\om(z) - \delta V'(z)) K(x,z)
\eeq

\bea
&& \delta W^{(g)}_{n+1}(x,L) \cr
&=& - \sum_i \Res_{z\to s_i} W^{(g)}_{n+1}(z,L) \,\,  \delta G(x,z) \cr
&& + \sum_i \Res_{z\to s_i} W^{(g)}_{n+1}(z,L) \,\, (2\delta\om(z) - \delta V'(z)) \, K(x,z) \cr
&& +2 \sum_i \Res_{z\to s_i}\sum_k \Res_{x''\to s_k}\,  K(x,z)\,\,  \delta V(x'')\,\,  B(z,x'') W_{n+1}^{(g)}(z,L)  \cr
&& + \, \sum_i \Res_{z\to s_i} K(x,z)\,\,  \delta V'(z)\,\,  W_{n+1}^{(g)}(z,L)  \cr
&& - \sum_k \Res_{x''\to s_k}\,  \delta V(x'')\,\,  W_{n+2}^{(g)}(z,L,x'')  \cr
\eea

We have:
\beq
 \sum_i \Res_{z\to s_i} W^{(g)}_{n+1}(z,L) \,\,  \delta G(x,z) =0
\eeq
because the integrand is a rational fraction, and we have taken the sum of residues at all poles.

Using \eq{varomResB}, we are thus left with:
\beq
\delta W^{(g)}_{n+1}(x,L) 
= - \sum_k \Res_{x''\to s_k}\,  \delta V(x'')\,\,  W_{n+2}^{(g)}(z,L,x'')  
\eeq
which proves the recursion hypothesis for $2g+n+1$.
QED.

\appendix{Proof of theorem \ref{thResY}}
\label{approofthResY}

{\bf Theorem \ref{thResY}}

For $k=0,1$, 
$W_n^{(g)}$ satify the equation:
\bea
&& \Big(- \sum_{i=1}^n x_i^k{\partial \over \partial x_i}\Big)\, W_n^{(g)}(x_1,\dots,x_n)  \cr
&=& \sum_i \Res_{x_{n+1}\to s_i}\,\, x_{n+1}^k\,V'(x_{n+1})\,\,W_{n+1}^{(g)}(x_1,\dots,x_n,x_{n+1})
\eea

\proof{

Since $W_{n+1}^{(g)}$ has poles only at the $s_i$'s we have (with as usual $J=\{x_1,\dots,x_n\}$):
\bea
&& \sum_i \Res_{x\to s_i}\,\, x^k\,V'(x)\,\,W_{n+1}^{(g)}(J,x)  \cr
&=& \sum_i \Res_{x\to s_i}\,\, x^k\,Y(x)\,\,W_{n+1}^{(g)}(J,x)  \cr
\eea
Then using theorem \ref{thWngPng}, we have:
\bea
&& \sum_i \Res_{x\to s_i}\,\, x^k\,V'(x)\,\,W_{n+1}^{(g)}(J,x)  \cr
&=& \sum_i \Res_{x\to s_i}\,\, x^k\,Y(x)\,\,W_{n+1}^{(g)}(J,x)  \cr
&=& \sum_i \Res_{x\to s_i}\,\, x^k\,
\Big[\hbar \partial_x W_{n+1}^{(g)}(J,x) + U_{n+1}^{(g)}(x,J) - P_{n+1}^{(g)}(x;J) 
-\sum_{j=1}^n \partial_{x_j}\, {W_n^{(g)}(J)\over x-x_j}\Big] \cr
&=& \sum_i \Res_{x\to s_i}\,\, x^k\,
\Big[\hbar \partial_x W_{n+1}^{(g)}(J,x) + U_{n+1}^{(g)}(x,J)  \Big] \cr
\eea
Notice that if $n\geq 1$, $W_{n+1}^{(g)}(J,x)$ behaves like $O(1/x^2)$ at $x\to\infty$, and thus, if $k\leq 1$, $x^k\,\partial_x W_{n+1}^{(g)}(J,x)$ behaves like $O(1/x^2)$. Since we take the residues at all poles, the sum of residues vanish and thus:
\bea
&& \sum_i \Res_{x\to s_i}\,\, x^k\,V'(x)\,\,W_{n+1}^{(g)}(J,x)  \cr
&=& \sum_i \Res_{x\to s_i}\,\, x^k\, U_{n+1}^{(g)}(x,J)  \cr
\eea
Notice that $U_{n+1}^{(g)}(x,J)$ (defined in \eq{defUng2}), behaves at most like $O(1/x^3)$ for large $x$, and thus, if $k\leq 1$, the product $x^k\,U_{n+1}^{(g)}(x,J)$ is a rational fraction, which behaves like $O(1/x^2)$ for large $x$. Its only poles can be at $x=s_i$ or at $x=x_j$. Therefore the sum of residues at $s_i$'s, can be replaced by the sum of residues at $x_j$'s:
\bea
&& \sum_i \Res_{x\to s_i}\,\, x^k\,V'(x)\,\,W_{n+1}^{(g)}(J,x)  \cr
&=& - \sum_{j=1}^n \Res_{x\to x_j}\,\, x^k\, U_{n+1}^{(g)}(x,J)  \cr
\eea
The only terms in $U_{n+1}^{(g)}(x,J)$ which have poles at $x=x_j$, are the terms containing a $B(x,x_j)$, i.e.:
\bea
 \sum_i \Res_{x\to s_i}\,\, x^k\,V'(x)\,\,W_{n+1}^{(g)}(J,x)  
&=& - 2\sum_{j=1}^n \Res_{x\to x_j}\,\, x^k\, B(x,x_j) \, W_n^{(g)}(x,J/\{x_j\})  \cr
&=& - \sum_{j=1}^n \Res_{x\to x_j}\,\, x^k\, {1\over (x-x_j)^2} \, W_n^{(g)}(x,J/\{x_j\})  \cr
&=& - \sum_{j=1}^n {\partial \over \partial x_j}\, \Big( x_j^k\, \, W_n^{(g)}(x_1,\dots,x_n)\, \Big)  \cr
\eea

}

\appendix{Proof of theorem \ref{thdilaton}}
\label{approofthdilaton}

{\bf Theorem \ref{thdilaton}:}

{\it
For $n\geq 1$, $W_n^{(g)}$ satify the equation:
\beq
(2-2g-n-\hbar {\partial\over \partial \hbar})\, \ovl{W}_{n}^{(g)}(x_1,\dots,x_n)
= - \sum_i \Res_{x_{n+1}\to s_i}\,\, V(x_{n+1})\,\,\ovl{W}_{n+1}^{(g)}(x_1,\dots,x_n,x_{n+1})
\eeq

}

\bigskip

\subsection{$\hbar$ derivatives for $w(z)$}
We have:
$$V'(s_i)=2\hbar \sum_{\neq i}\frac{1}{s_i-s_j}$$
Taking the derivative with respect to $\hbar$ gives:
$$\hbar  V''(s_i)\partial_\hbar s_i=V'(s_i)-2\hbar^2\sum_{j \neq i} \frac{\partial_\hbar si-\partial_\hbar s_j}{(s_i-s_j)^2}$$
and so
$$V'(s_i)=\hbar\left( V''(s_i)\partial_\hbar s_i+2\hbar\sum_{j \neq i} \frac{\partial_\hbar si-\partial_\hbar s_j}{(s_i-s_j)^2}\right)$$
We recognize the general term of the matrix $T$ and find:
$$V'(s_i)=\hbar^2\sum_{j}T_{i,j}\partial_\hbar s_j$$
Multiplying by the matrix $A$ gives:
\beq \encadremath{\hbar^2 \partial_{\hbar}s_i=\sum_j A_{i,j}V'(s_j)}\eeq
We can use this result to compute:
\bea
\hbar \partial_\hbar \om(x)
&=&\om(x)+\hbar^2\sum_i \frac{\partial_\hbar si}{(x-s_i)^2}\cr
&=& \om(x)+\sum_{i,j}\frac{A_{i,j}V'(s_j)}{(x-s_i)^2} \cr
&=& \om(x)+\sum_k \Res_{x' \to s_k}\sum_{i,j}\frac{A_{i,j}V'(x')}{(x-s_i)^2(x'-s_j)}\cr
&=& \om(x)+\sum_k \Res_{x' \to s_k}\sum_{i,j}\frac{A_{i,j}V(x')}{(x-s_i)^2(x'-s_j)^2}\cr
&=& \om(x)+\sum_k \Res_{x' \to s_k}\ovl{W}_2^{(0)}(x,x')V(x') \cr
&=& \om(x)+\sum_k \Res_{x' \to s_k}W_2^{(0)}(x,x')V(x')\cr
\eea
Thus we have proved the case $n=1, g=0$ of the theorem:
\beq \label{Derivomhbar} \encadremath{\hbar \partial_\hbar \om(x)=\om(x)+\sum_k \Res_{x' \to s_k}W_2^{(0)}(x,x')V(x')}\eeq

\subsection{$\hbar$ derivatives for $W_2^{(0)}(z)$}

We have seen in appendix \ref{approofthvariationV}, \eq{apploopeqW20}, that $\ovl{W}_2^{(0)}(x,x')$ satisfies the loop equation:
\beq
(2\om(x)-V'(x)+\hbar {\partial_x}) \,\, \ovl{W}_2^{(0)}(x,x') + {\partial \over \partial x'} {\om(x)-\om(x')\over x-x'} = - P_2^{(0)}(x,x')
\eeq
where $P_2^{(0)}(x,x')$ has no pole at  $x\to s_i$'s.

\medskip
Then we take the derivation $\hbar \partial_\hbar$ of this equation:
\bea
(2\om(x)-V'(x)+\hbar \partial_x) \,\, \hbar \partial_\hbar \ovl{W}_2^{(0)}(x,x')+\hbar \partial_x \ovl{W}_2^{(0)}(x,x')+2\hbar \partial_\hbar w(x)\ovl{W}_2^{(0)}(x,x') \cr
=
- {\partial \over \partial x'} {\hbar\partial_\hbar \om(x)-\hbar\partial_\hbar \om(x')\over x-x'}  - \hbar\partial_\hbar P_2^{(0)}(x,x') \cr
\eea
$\hbar \partial_\hbar \ovl{W}_2^{(0)}(x,x')$ is a rational fraction of $x$, with poles only at the $s_i$'s, and $\hbar\partial_\hbar P_2^{(0)}(x,x')$ has no pole at  $x\to s_i$'s.
We thus write:
\bea
&& \hbar \partial_\hbar W_2^{(0)}(x,x')  \cr
&=&\hbar \partial_\hbar \ovl{W}_2^{(0)}(x,x')\cr
&=& \Res_{z\to x} G(x,z)\,\, \hbar \partial_\hbar \ovl{W}_2^{(0)}(z,x') \cr
&=& - \sum_i \Res_{z\to s_i} G(x,z)\,\, \hbar \partial_\hbar \ovl{W}_2^{(0)}(z,x') \cr
&=& - \sum_i \Res_{z\to s_i} \left( (2\om(z)-V'(z)-\hbar {\partial_z})K(x,z)\right) \,\, \hbar \partial_\hbar \ovl{W}_2^{(0)}(z,x') \cr
&=& - \sum_i \Res_{z\to s_i} K(x,z)\,\, \left( (2\om(z)-V'(z)+\hbar {\partial_z}) \,\, \hbar \partial_\hbar \ovl{W}_2^{(0)}(z,x') \right)\cr
&=&  \sum_i \Res_{z\to s_i} K(x,z)\,\, \Big(  (2\hbar \partial_\hbar \om(z)) \,\, \ovl{W}_2^{(0)}(z,x') \cr
&& + {\partial \over \partial x'} {\hbar \partial_\hbar\om(z)+\hbar \partial_\hbar\om(x')\over z-x'}  + \hbar \partial_\hbar P_2^{(0)}(z,x')
+\hbar \partial_z \ovl{W}_2^{(0)}(z,x') \Big)\cr
&=&  \sum_i \Res_{z\to s_i} K(x,z)\,\, \Big(    2\ovl{W}_2^{(0)}(z,x')\,\,\hbar \partial_\hbar \om(z) 
 + {\hbar \partial_\hbar \om(z)\over (z-x')^2}+\hbar \partial_z \ovl{W}_2^{(0)}(z,x')   \Big)\cr
&=&  \sum_i \Res_{z\to s_i} K(x,z)\,\,\Big(2W_2^{(0)}(z,x')\,\hbar \partial_\hbar \om(z)+\hbar \partial_z W_2^{(0)}(z,x')\Big)\cr
\eea
Then, we use \eq{Derivomhbar}, and we get:
\bea
&& \hbar \partial_\hbar W_2^{(0)}(x,x') \cr
&=& \sum_i \Res_{z\to s_i} K(x,z)\,\,\Big(2W_2^{(0)}(z,x')w(z)+\hbar \partial_z W_2^{(0)}(z,x')\Big)\cr
&& + 2\sum_{i,k} \Res_{z\to s_i}\Res_{x'' \to s_k} K(x,z)W_2^{(0)}(z,x')W_2^{(0)}(z,x'')V(x'')\cr
&=& \sum_i \Res_{z\to s_i} W_2^{(0)}(z,x')\,\,\Big(2w(z)-\hbar \partial_z \Big)K(x,z)\cr
&& +\sum_{i,k} \Res_{z\to s_i}\Res_{x'' \to s_k} K(x,z)W_2^{(0)}(z,x')G(z,x'')V'(x'')\cr
&=& \sum_i \Res_{z\to s_i} W_2^{(0)}(z,x')\,\,(G(x,z)+V'(z)K(x,z))\cr
&& +\sum_{i,k} \Res_{z\to s_i}\Res_{x'' \to s_k} K(x,z)W_2^{(0)}(z,x')G(z,x'')V'(x'')\cr
&=& \sum_i \Res_{z\to s_i} W_2^{(0)}(z,x')\,\,G(x,z)\cr
&& +\sum_{i,k} \Res_{z\to s_i}\Res_{x'' \to s_k} K(x,z)W_2^{(0)}(z,x')G(z,x'')V'(x'')\cr
&& +\sum_{i} \Res_{z\to s_i}\Res_{x'' \to z} K(x,z)W_2^{(0)}(z,x')G(z,x'')V'(x'')\cr
&=& \sum_i \Res_{z\to s_i} W_2^{(0)}(z,x')\,\,G(x,z)\cr
&& +\sum_{i,k} \Res_{x'' \to s_k}\,\Res_{z\to s_i} K(x,z)W_2^{(0)}(z,x')G(z,x'')V'(x'')\cr
&=& \sum_i \Res_{z\to s_i} W_2^{(0)}(z,x')\,\,G(x,z)\cr
&& +2 \sum_{i,k} \Res_{x'' \to s_k}\,\Res_{z\to s_i} K(x,z)W_2^{(0)}(z,x')B(z,x'')V(x'')\cr
&=& \sum_i \Res_{z\to s_i} B(z,x')\,\,G(x,z)\cr
&& + \sum_{k} \Res_{x'' \to s_k}\,W_3^{(0)}(x,x',x'')V(x'')\cr
\eea
We now use the fact that $G(x,z)$ and $B(z,x')$ are rational fractions whose only poles are $s_i$'s, as well as $z=x$ and $z=x'$, and we write:
\bea
&& \sum_i \Res_{z\to s_i} B(z,x')\,\,G(x,z)\cr
&=& - \Res_{z\to x} B(z,x')\,\,G(x,z) - \Res_{z\to x'} B(z,x')\,\,G(x,z)\cr
&=& - \Res_{z\to x} B(z,x')\,\,{1\over z-x} - {1\over 2}\,\Res_{z\to x'} {1\over (z-x')^2}\,\,G(x,z)\cr
&=& - \Res_{z\to x} B(z,x')\,\,{1\over z-x} +\,\Res_{z\to x'} {1\over z-x'}\,\,B(x,z)\cr
&=& - B(x,x')+B(x,x')\cr
&=& 0
\eea

So that eventually we have proved the case $n=2,g=0$ of the theorem:
\beq\label{varBResW3hbar}
\encadremath{
\hbar \partial_\hbar W_2^{(0)}(x,x') =  \sum_k \Res_{x''\to s_k} W_3^{(0)}(x,x',x'')\,\, V(x'')    
}\eeq

\subsection{Recursion for higher correlators}

We proceed by recursion on $2g+n$.

From theorem \ref{thWngPng}, we have that:
\bea
&& (Y(x)-\hbar \partial_x) \hbar\partial_{\hbar} W_{n+1}^{(g)}(x,L) \cr
&=& \hbar\partial_{\hbar} U_{n+1}^{(g)}(x;L) + \hbar \partial_x W_{n+1}^{(g)}(x,L)
- W_{n+1}^{(g)}(x,L) \,  \hbar\partial_{\hbar} Y(x) \cr
&& -  \hbar\partial_{\hbar} \left( P_{n+1}^{(g)}(x;L) +\sum_{x_j\in L} {\partial \over \partial x_j}\, { {\ovl{W}}_{n}^{(g)}(L)\over x-x_j} \right)
\eea
where the term on the last line has no pole at $x=s_i$.
This implies that:
\bea
&& \sum_i \Res_{x\to s_i} K(x_0,x)\,\Big((Y(x)-\hbar \partial_x) \hbar\partial_{\hbar} W_{n+1}^{(g)}(x,L)\Big) \cr
&=& \sum_i \Res_{x\to s_i} K(x_0,x)\,\Big( \hbar\partial_{\hbar} U_{n+1}^{(g)}(x;L) + \hbar \partial_x W_{n+1}^{(g)}(x,L) \cr
&& - W_{n+1}^{(g)}(x,L) \,  \hbar\partial_{\hbar} Y(x)  \Big) 
\eea
We have:
\bea
&& \sum_i \Res_{x\to s_i} K(x_0,x)\,\Big((Y(x)-\hbar \partial_x) \hbar\partial_{\hbar} W_{n+1}^{(g)}(x,L)\Big) \cr
&=& \sum_i \Res_{x\to s_i} \hbar\partial_{\hbar} W_{n+1}^{(g)}(x,L)\,(Y(x)+\hbar \partial_x) K(x_0,x)  \cr
&=& - \sum_i \Res_{x\to s_i} \hbar\partial_{\hbar} W_{n+1}^{(g)}(x,L)\,G(x_0,x)  \cr
&=&  \Res_{x\to x_0} \hbar\partial_{\hbar} W_{n+1}^{(g)}(x,L)\,G(x_0,x)  \cr
&=&   \hbar\partial_{\hbar} W_{n+1}^{(g)}(x_0,L)  
\eea
and therefore:
\bea
&&   \hbar\partial_{\hbar} W_{n+1}^{(g)}(x_0,L)  \cr
&=& \sum_i \Res_{x\to s_i} K(x_0,x)\,\Big( \hbar\partial_{\hbar} U_{n+1}^{(g)}(x;L) + \hbar \partial_x W_{n+1}^{(g)}(x,L)
- W_{n+1}^{(g)}(x,L) \,  \hbar\partial_{\hbar} Y(x)  \Big) \cr
\eea
From the recursion hypothesis we have:
\bea
&& \hbar\partial_{\hbar} U_{n+1}^{(g)}(x;L) \cr
&=& \hbar\partial_{\hbar} W_{n+2}^{(g-1)}(x,x,L)
+ \sum_{k=0}^g\sum'_{J\subset L} W_{1+|J|}^{(k)}(x,J) \hbar\partial_{\hbar} W_{1+n-|J|}^{(g-k)}(x,L/J) \cr
&& + \sum_{k=0}^g\sum'_{J\subset L}  W_{1+n-|J|}^{(g-k)}(x,L/J) \hbar\partial_{\hbar}  W_{1+|J|}^{(k)}(x,J)\cr
&=& (2-2(g-1)-(n+2))  W_{n+2}^{(g-1)}(x,x,L) + \sum_i\Res_{x'\to s_i}  W_{n+3}^{(g-1)}(x,x,L,x')\, V(x') \cr
&& + \sum_{k=0}^g\sum'_{J\subset L} (2-2(g-k)-(1+n-|J|))\,W_{1+|J|}^{(k)}(x,J) \, W_{1+n-|J|}^{(g-k)}(x,L/J) \cr
&& + \sum_{k=0}^g\sum'_{J\subset L}  (2-2k-(1+|J|))\, W_{1+n-|J|}^{(g-k)}(x,L/J) \,  W_{1+|J|}^{(k)}(x,J)\cr
&& + \sum_i \Res_{x'\to s_i} V(x')\sum_{k=0}^g\sum'_{J\subset L} W_{2+|J|}^{(k)}(x,J,x') \, W_{1+n-|J|}^{(g-k)}(x,L/J) \cr
&& + \sum_i \Res_{x'\to s_i} V(x')\sum_{k=0}^g\sum'_{J\subset L} W_{1+|J|}^{(k)}(x,J) \, W_{2+n-|J|}^{(g-k)}(x,L/J,x') \cr
&=& (2-2g-n)\,\,  U_{n+1}^{(g)}(x;L) \cr 
&& + \sum_i \Res_{x'\to s_i} V(x')\,(U_{n+2}^{(g)}(x;x',L) -2B(x,x') W_{n+1}^{(g)}(x,L) )  
\eea

Thus we have:
\bea
&&   \hbar\partial_{\hbar} W_{n+1}^{(g)}(x_0,L)  \cr
&=& (2-2g-n)\sum_i \Res_{x\to s_i} K(x_0,x)\,U_{n+1}^{(g)}(x;L) \cr
&& + \sum_i \Res_{x\to s_i} K(x_0,x)  \sum_j \Res_{x'\to s_j} V(x')\, (U_{n+2}^{(g)}(x;x',L)-2B(x,x') W_{n+1}^{(g)}(x,L) ) \cr
&& + \sum_i \Res_{x\to s_i} K(x_0,x)\,\Big(  \hbar \partial_x W_{n+1}^{(g)}(x,L)
- W_{n+1}^{(g)}(x,L) \,  \hbar\partial_{\hbar} Y(x)  \Big) \cr
&=& (2-2g-n) W_{n+1}^{(g)}(x_0,L) \cr
&& + \sum_j \Res_{x'\to s_j}  \sum_i \Res_{x\to s_i} K(x_0,x)  V(x')\, (U_{n+2}^{(g)}(x;x',L)-2B(x,x') W_{n+1}^{(g)}(x,L) ) \cr
&& + \sum_i \Res_{x\to s_i} K(x_0,x)\,\Big(  \hbar \partial_x W_{n+1}^{(g)}(x,L)
- W_{n+1}^{(g)}(x,L) \,  \hbar\partial_{\hbar} Y(x)  \Big) \cr
&=& (2-2g-n) W_{n+1}^{(g)}(x_0,L)  + \sum_j \Res_{x'\to s_j}   V(x')\, W_{n+2}^{(g)}(x_0,x',L) \cr
&& -2 \sum_j \Res_{x'\to s_j}  \sum_i \Res_{x\to s_i} K(x_0,x)  V(x')\, B(x,x') W_{n+1}^{(g)}(x,L)  \cr
&& + \sum_i \Res_{x\to s_i} K(x_0,x)\,\Big(  \hbar \partial_x W_{n+1}^{(g)}(x,L)
- W_{n+1}^{(g)}(x,L) \,  \hbar\partial_{\hbar} Y(x)  \Big) \cr
&=& (2-2g-n) W_{n+1}^{(g)}(x_0,L)  + \sum_j \Res_{x'\to s_j}   V(x')\, W_{n+2}^{(g)}(x_0,x',L) \cr
&& -2 \sum_i \Res_{x\to s_i} \sum_j \Res_{x'\to s_j}   K(x_0,x)  V(x')\, B(x,x') W_{n+1}^{(g)}(x,L)  \cr
&& -2\sum_i \Res_{x\to s_i} \Res_{x'\to x}   K(x_0,x)  V(x')\, B(x,x') W_{n+1}^{(g)}(x,L)  \cr
&& + \sum_i \Res_{x\to s_i} K(x_0,x)\,\Big(  \hbar \partial_x W_{n+1}^{(g)}(x,L)
- W_{n+1}^{(g)}(x,L) \,  \hbar\partial_{\hbar} Y(x)  \Big) 
\eea

Notice that:
\beq
 \hbar\partial_{\hbar} Y(x)
+ 2\sum_j \Res_{x'\to s_j} B(x,x') V(x')
+ 2\Res_{x'\to x} B(x,x') V(x')
= Y(x)
\eeq
therefore:
\bea
&&   \hbar\partial_{\hbar} W_{n+1}^{(g)}(x_0,L)  \cr
&=& (2-2g-n) W_{n+1}^{(g)}(x_0,L)  + \sum_j \Res_{x'\to s_j}   V(x')\, W_{n+2}^{(g)}(x_0,x',L) \cr
&& + \sum_i \Res_{x\to s_i} K(x_0,x)\,\Big(  \hbar \partial_x W_{n+1}^{(g)}(x,L)
- Y(x) W_{n+1}^{(g)}(x,L)   \Big) \cr
&=& (2-2g-n) W_{n+1}^{(g)}(x_0,L)  + \sum_j \Res_{x'\to s_j}   V(x')\, W_{n+2}^{(g)}(x_0,x',L) \cr
&& - \sum_i \Res_{x\to s_i} W_{n+1}^{(g)}(x,L)\,(  Y(x)+\hbar \partial_x )K(x_0,x)  \cr
&=& (2-2g-n) W_{n+1}^{(g)}(x_0,L)  + \sum_j \Res_{x'\to s_j}   V(x')\, W_{n+2}^{(g)}(x_0,x',L) \cr
&& + \sum_i \Res_{x\to s_i} W_{n+1}^{(g)}(x,L)\,G(x_0,x)  \cr
&=& (2-2g-n) W_{n+1}^{(g)}(x_0,L)  + \sum_j \Res_{x'\to s_j}   V(x')\, W_{n+2}^{(g)}(x_0,x',L) \cr
&& -  \Res_{x\to x_0} W_{n+1}^{(g)}(x,L)\,G(x_0,x)  \cr
&=& (2-2g-n) W_{n+1}^{(g)}(x_0,L)  + \sum_j \Res_{x'\to s_j}   V(x')\, W_{n+2}^{(g)}(x_0,x',L) \cr
&& - W_{n+1}^{(g)}(x_0,L)  \cr
&=& (2-2g-n-1) W_{n+1}^{(g)}(x_0,L)  + \sum_j \Res_{x'\to s_j}   V(x')\, W_{n+2}^{(g)}(x_0,x',L) 
\eea
i.e. we have proved the theorem for $2g+n+1$.

\appendix{Free Energies}
\label{approofFg}

Here we consider $g\geq 2$.

\medskip

The free energies defined in \eq{defFg}, automatically satisfy theorem \ref{thdilaton}, and thus are homogeneous:
\beq
F^{(g)}(\l V,\l \hbar) = \l^{2-2g}\,\, F^{(g)}( V,\hbar)
\eeq

Here we show that they satisfy theorem \ref{thvariationV}.

\medskip

We start from the definition:
\beq\label{defFg1}
F^{(g)} =  \hbar^{2-2g}\,\int_0^{\hbar}\, {d\td{\hbar}\over {\td\hbar}^{3-2g}}\,\, \sum_i \Res_{x\to s_i}\,\, V(x)\,\,\, \left. W_{1}^{(g)}(x)\right|_{{\td\hbar}}
\eeq
and we compute the loop operator applied to $F^{(g)}$:
\bea
\delta_{x_1}\, F^{(g)} 
&=&  \hbar^{2-2g}\,\int_0^{\hbar}\, {d\td{\hbar}\over {\td\hbar}^{3-2g}}\,\, \sum_i \Res_{x\to s_i}\left(\,\, V(x)\,\,\,  W_{2}^{(g)}(x,x_1)+ \delta_{x_1}\,V(x)\,\,\,  W_{1}^{(g)}(x) \right)_{{\td\hbar}}  \cr
&=&  \hbar^{2-2g}\,\int_0^{\hbar}\, {d\td{\hbar}\over {\td\hbar}^{3-2g}}\,\, \sum_i\Res_{x\to s_i}\,\left( \,  V(x)\,\,\,  W_{2}^{(g)}(x,x_1) + {W_{1}^{(g)}(x)\over x-x_1} \right)_{{\td\hbar}}  \cr
&=&  \hbar^{2-2g}\,\int_0^{\hbar}\, {d\td{\hbar}\over {\td\hbar}^{3-2g}}\,\, \left(   \left(\sum_i\Res_{x\to s_i} V(x)\,\,\,  W_{2}^{(g)}(x,x_1) \right) -  W_{1}^{(g)}(x_1)\right)_{{\td\hbar}}  \cr
&=&  \hbar^{2-2g}\,\int_0^{\hbar}\, {d\td{\hbar}\over {\td\hbar}^{3-2g}}\,\, \left(  \td{h}^{2-2g}\,{d (\td{h}^{2g-1}\,W_{1}^{(g)}(x_1))\over d\td{h}} -W_{1}^{(g)}(x_1) \right)_{{\td\hbar}}  \cr
&=&  \hbar^{2-2g}\,\int_0^{\hbar}\, \left({1\over \td{h}}\,d \left(\td{h}^{2g-1}\,W_{1}^{(g)}(x_1)\right) - {d\td{\hbar}\over {\td\hbar}^{3-2g}}\,\,W_{1}^{(g)}(x_1) \right)_{{\td\hbar}} \cr
\eea
we integrate by parts, and since $2g-2>0$, there is no boundary term coming from the bound at $0$, and thus:
\bea
\delta_{x_1}\, F^{(g)} 
&=& W_1^{(g)}(x_1) + \hbar^{2-2g}\,\int_0^{\hbar}\, \left( \td{h}^{2g-3}\,W_{1}^{(g)}(x_1)- {\td\hbar}^{2g-3}\,\,W_{1}^{(g)}(x_1) \right)_{{\td\hbar}} \, d\td{h}\cr
&=& W_1^{(g)}(x_1)
\eea
Therefore we have proved that the loop operator acting on $F^{(g)}$ is indeed $W_1^{(g)}$, i.e. we have proved theorem \ref{thvariationV}.

\appendix{$F^{(0)}$}
\label{approofF0}

We have defined $F^{(0)}$ as:
\beq
F^{(0)} =  -\hbar \sum_i  V(s_i) + \hbar^2 \sum_{i\neq j} \ln{(s_i-s_j)}
\eeq

\bigskip
$\bullet$ Proof of theorem \ref{thvariationV} for $F^{(0)}$:

consider a variation $\delta V$, we have: 
\bea 
\delta F^{(0)}
&=& -\hbar \sum_i \delta V(s_i) - \hbar \sum_i V'(s_i) \delta s_i + 2\hbar^2 \sum_{j\neq i} {\delta s_i\over s_i-s_j} \cr
&=& -\hbar \sum_i \delta V(s_i)  \cr
&=& - \sum_i \Res_{x\to s_i} \om(x)\,  \delta V(x)  
\eea

\bigskip
$\bullet$ Proof of theorem \ref{thdilaton} for $F^{(0)}$:

we have:
\bea
\hbar \partial_{\hbar} F^{(0)}
&=& -\hbar \sum_i  V(s_i) + 2\hbar^2 \sum_{i\neq j} \ln{(s_i-s_j)} \cr
&& -\hbar^2 \sum_i {\partial s_i\over \partial \hbar}\,\left(V'(s_i) - 2\hbar\sum_{j\neq i} {1\over s_i-s_j} \right) \cr
&=& -\hbar \sum_i  V(s_i) + 2\hbar^2 \sum_{i\neq j} \ln{(s_i-s_j)} \cr
&=& 2 F^{(0)} + \hbar \sum_i  V(s_i)  \cr
&=& 2 F^{(0)} +  \sum_i \Res_{x\to s_i} \om(x)\, V(x)  
\eea
Therefore:
\beq (2-\hbar\partial_\hbar)F_0=-\sum_i \Res_{x \to s_i}V(x)w(x)\eeq

\appendix{$F^{(1)}$}
\label{approofF1}

We have defined $F^{(1)}$ as:
\bea
F^{(1)} 
&=& {1\over 2}\,\ln{(\det A)}\, +{F^{(0)}\over \hbar^2} + \ln{(\Delta(s)^2)} \cr
&=& {1\over 2}\,\ln{(\det A)}\, - {1\over \hbar} \sum_i V(s_i) + \sum_{i\neq j} \ln{(s_i-s_j)} + \sum_{i\neq j} \ln{(s_i-s_j)} \cr
&=& {1\over 2}\,\ln{(\det A)}\, - {1\over \hbar} \sum_i V(s_i) + 2\sum_{i\neq j} \ln{(s_i-s_j)}  
\eea

\bigskip
$\bullet$ Proof of theorem \ref{thvariationV} for $F^{(1)}$:

Let us start from $W_1^{(1)}$
\bea
W_1^{(1)}(x) 
&=& \sum_i \Res_{z\to s_i} K(x,z)\,\ovl{W}_2(z,z) \cr
&=& \sum_i \Res_{z\to s_i} K(x,z)\,\Big[{A_{i,i}\over (z-s_i)^4} + 2\sum_{j\neq i} {A_{i,j}\over (z-s_i)^2(z-s_j)^2} \Big] \cr
&=& \sum_i \Res_{z\to s_i} K(x,z)\,{A_{i,i}\over (z-s_i)^4}  \cr
&& +2 \sum_i\sum_{j\neq i}  K'(x,s_i)\, {A_{i,j}\over (s_i-s_j)^2}  \cr
&& -4 \sum_i\sum_{j\neq i}  K(x,s_i)\, {A_{i,j}\over (s_i-s_j)^3}  \cr
\eea

We have:
\bea
&& \sum_i \Res_{z\to s_i} K(x,z)\,{A_{i,i}\over (z-s_i)^4} \cr
&=& {1\over 3} \sum_i \Res_{z\to s_i} K'(x,z)\,{A_{i,i}\over (z-s_i)^3} \cr
&=& {1\over 3} \sum_i \Res_{z\to s_i} ({2\over z-s_i} + 2 \om_i(z) - {1\over \hbar}V'(z))K(x,z) \,{A_{i,i}\over (z-s_i)^3} \cr
&& - {1\over 3\hbar } \sum_i \Res_{z\to s_i} G(x,z) \,{A_{i,i}\over (z-s_i)^3} 
\eea
Therefore:
\bea
&& \sum_i \Res_{z\to s_i} K(x,z)\,{A_{i,i}\over (z-s_i)^4} \cr
&=&  \sum_i \Res_{z\to s_i} ( 2 \om_i(z) - {1\over \hbar}V'(z))K(x,z) \,{A_{i,i}\over (z-s_i)^3} \cr
&& -  {1\over \hbar}\,\sum_i \Res_{z\to s_i} G(x,z) \,{A_{i,i}\over (z-s_i)^3} \cr
&=&  \sum_i \Res_{z\to s_i} \Big[{ 2 \om_i(z) - {1\over \hbar}V'(z)\over z-s_i}\,K(x,z)\Big] \,{A_{i,i}\over (z-s_i)^2} \cr
&& -  {1\over 2\hbar}\,\sum_i \Res_{z\to s_i} G'(x,z) \,{A_{i,i}\over (z-s_i)^2} \cr
&=&  \sum_i A_{i,i}\, \Big[{ 2 \om_i(z) - {1\over \hbar}V'(z)\over z-s_i}\,K(x,z)\Big]'_{z= s_i}  \cr
&& + {1\over \hbar}\,\sum_i \Res_{z\to s_i} B(x,z) \,{A_{i,i}\over (z-s_i)^2} \cr
&=& {1\over 2} \sum_i  ( 2 \om_i''(s_i) - {1\over \hbar}V'''(s_i))K(x,s_i) \,{A_{i,i}} \cr
&&  - \sum_i  K'(x,s_i) \,{A_{i,i} T_{i,i}} \cr
&& +  {1\over \hbar}\,\sum_i \Res_{z\to s_i} B(x,z) \,{A_{i,i}\over (z-s_i)^2} 
\eea

Notice that:
\bea
\Res_{x\to s} K(x,s_i) \delta V(x) 
&=& {1\over \hbar}\, \sum_{j} \Res_{x\to s} {A_{i,j} \delta V(x) \over (x-s_j)^2} \cr
&=& {1\over \hbar}\,\sum_{j}  A_{i,j} \delta V'(s_j) \cr
&=& - \delta s_i 
\eea

\bea
\Res_{x\to s} K'(x,s_i) \delta V(x) 
&=& -{1\over \hbar}\,\sum_j \delta_{i,j}\delta V(s_j) - 2\sum_{j\neq i} {\delta s_j\over s_i-s_j}
\eea

\bea
\Res_{x\to s} \Res_{z\to s_i} {B(x,z)\over (z-s_i)^2}\,\, \delta V(x) 
&=&  \Res_{z\to s_i} \Res_{x\to s}{B(x,z)\over (z-s_i)^2}\,\, \delta V(x)  \cr
&& +  \Res_{z\to s_i} \Res_{x\to z} {B(x,z)\over (z-s_i)^2}\,\, \delta V(x)  \cr
&=&  \Res_{z\to s_i} \Res_{x\to s}{A_{j,l}\over (x-s_l)^2(z-s_j)^2 (z-s_i)^2}\,\, \delta V(x)  \cr
&& +  {1\over 2}\Res_{z\to s_i} \Res_{x\to z} {1\over (x-z)^2(z-s_i)^2}\,\, \delta V(x)  \cr
&=&  {\hbar}\,\Res_{z\to s_i} \Res_{x\to s}{K(x,s_j)\over (z-s_j)^2 (z-s_i)^2}\,\, \delta V(x)  \cr
&& +  {1\over 2}\Res_{z\to s_i}  {1\over (z-s_i)^2}\,\, \delta V'(z)  \cr
&=& - \hbar \, \Res_{z\to s_i} {\delta s_j\over (z-s_j)^2 (z-s_i)^2}  +  {1\over 2}\,\, \delta V''(s_i)  \cr
&=& 2\hbar\,  {\delta s_j\over (s_i-s_j)^3 }  +  {1\over 2}\,\, \delta V''(s_i)  
\eea

That gives:
\bea
&& \Res_x \Res_{z\to s_i} {K(x,z)\, A_{i,i}\over (x-s_i)^4}\, \delta V(x) \cr
&=& - {1\over 2}   ( 2 \om_i''(s_i) - {1\over \hbar}V'''(s_i)) \delta s_i \,{A_{i,i}}  
 + {1\over \hbar}\, \sum_j \delta_{i,j}\delta V(s_j) \,{A_{i,i} T_{i,i}}   \cr
 && + 2\sum_{j\neq i} {\delta s_j\over s_i-s_j}    \,{A_{i,i} T_{i,i}} 
 +  2\hbar\,  {\delta s_j\over (s_i-s_j)^3 }\,A_{i,i}   +  {1\over 2}\,\, \delta V''(s_i) \,A_{i,i} \cr
&=& {1\over 2}  \delta (T_{i,i})  \,{A_{i,i}} 
  + {1\over \hbar}\, \sum_j \delta_{i,j}\delta V(s_j) \,{A_{i,i} T_{i,i}} 
  + 2\sum_{j\neq i} {\delta s_j\over s_i-s_j}    \,{A_{i,i} T_{i,i}} 
\eea

and thus:
\bea
&& \Res_{x\to s} W_1^{(1)}(x) \delta V(x) \cr
&=& \sum_i {1\over 2}  \delta (T_{i,i})  \,{A_{i,i}} 
  + {1\over \hbar}\, \sum_i\sum_j \delta_{i,j}\delta V(s_j) \,{A_{i,i} T_{i,i}} 
  + 2\sum_{j\neq i} {\delta s_j\over s_i-s_j}    \,{A_{i,i} T_{i,i}} \cr
&& -2 \sum_i\sum_{j\neq i}  {{1\over \hbar}\,\sum_l \delta_{i,l}\delta V(s_l)\, \over (s_i-s_j)^2} \,A_{i,j} 
 -4 \sum_i\sum_{j\neq i}\sum_{l\neq i}  {\delta s_l\, \over (s_i-s_l)(s_i-s_j)^2} \,A_{i,j} \cr
&& +4 \sum_i\sum_{j\neq i} { \delta s_i \over (s_i-s_j)^3} \,A_{i,j} \cr
&=& \sum_i {1\over 2}  \delta (T_{i,i})  \,{A_{i,i}} 
  + {1\over \hbar}\, \sum_i\sum_j \sum_l \delta_{i,j}\delta V(s_j) \,{A_{i,l} T_{l,i}} 
  + 2\sum_{j\neq i} {\delta s_j\over s_i-s_j}    \,{A_{i,i} T_{i,i}} \cr
&&  -4 \sum_i\sum_{j\neq i}\sum_{l\neq i}  {\delta s_l\, \over (s_i-s_l)(s_i-s_j)^2} \,A_{i,j} 
 +4 \sum_i\sum_{j\neq i} { \delta s_i \over (s_i-s_j)^3} \,A_{i,j} \cr
&=&  {1\over 2} \Tr A\,\, \delta T 
  + {1\over \hbar}\, \sum_i\sum_j \sum_l \delta_{i,j}\delta V(s_j) \,{A_{i,l} T_{l,i}} 
  + 2\sum_{j\neq i} {\delta s_j\over s_i-s_j}   \cr
&&  +  4\sum_{j\neq i}\sum_{l\neq i} {\delta s_j\over (s_i-s_j)(s_i-s_l)^2}    \,A_{i,l} 
 -4 \sum_{i\neq j\neq l}  {\delta s_l\, \over (s_i-s_l)(s_i-s_j)^2} \,A_{i,j} \cr
&=&  {1\over 2} \Tr A\,\, \delta T 
  + {1\over \hbar}\, \sum_j \delta V(s_j)  
  - \sum_{j\neq i} {\delta s_i-\delta s_j\over s_i-s_j}   \cr
&=&{1\over 2}\delta \ln{\det{T}}+ {1\over \hbar}\, \sum_j\delta(V(s_j)) 
- {1\over \hbar}\, \sum_j V'(s_j)\delta s_j- \sum_{j\neq i} {\delta s_i-\delta s_j\over s_i-s_j}   \cr
&=&{1\over 2}\delta \ln{\det{T}}+ {1\over \hbar}\, \sum_j\delta(V(s_j)) 
- 2\, \sum_j\sum_{i \neq j} {{\delta s_j}\over{s_j-s_i}}- \sum_{j\neq i} {\delta s_i-\delta s_j\over s_i-s_j}   \cr
&=&{1\over 2}\delta \ln{\det{T}}+ {1\over \hbar}\, \sum_j\delta(V(s_j)) 
- \, \sum_j\sum_{i \neq j} {{\delta s_j-\delta s_i}\over{s_j-s_i}}- \sum_{j\neq i} {\delta s_i-\delta s_j\over s_i-s_j}   \cr
&=&{1\over 2}\delta \ln{\det{T}}+ {1\over \hbar}\, \sum_j\delta(V(s_j))-2 \, \sum_{i \neq j} {{\delta s_j-\delta s_i}\over{s_j-s_i}} 
\eea

That implies:
\bea
F_1 
&=& -{1\over 2}\ln{\det{T}}   - {1\over \hbar}\, \sum_j V(s_j) +2 \,\sum_{i \neq j} \ln(s_i-s_j)
\eea

\beq
\encadremath{
F_1  = {1\over 2}\ln{\det{A}}  - {1\over \hbar}\, \sum_j V(s_j) +2 \,\sum_{i \neq j} \ln(s_i-s_j)
}\eeq

\appendix{Example $m=1$}
\label{appm1}

We choose $s=0$, and $V'(s) = v_2 s + v_3 s^2 + \sum v_{k+1} s^k$.

We have
\beq
\om(x) = {\hbar\over x}
\eeq
\beq
A= {\hbar\over v_2}
\eeq

\beq
K(x_1,x) = \sum_k K_k(x_1)\,\, x^k
\eeq
\beq
K_0 = {1\over v_2 x_1^2}
\virg
K_1=K_2=0
\eeq
\beq
K_3 = {1\over \hbar x_1^3} - {v_3 \over \hbar v_2 x_1^2}
\eeq

\beq
B(x_1,x_2) = {1\over 2(x_1-x_2)^2} + {A\over x_1^2 x_2^2}
\eeq

\bea
W_3^{(0)} 
&=& {2\hbar\over v_2^2\,x_1^2\,x_2^2\,x_3^2}\,({1\over x_1}+{1\over x_2}+{1\over x_3}) - {2\hbar\, v_3\over v_2^3\,x_1^2\,x_2^2\,x_3^2}
\eea

\bea
W_4^{(0)} 
&=& {6\hbar\over v_2^3\,x_1^2\,x_2^2\,x_3^2}\,({1\over x_1^2}+{1\over x_2^2}+{1\over x_3^2}+{1\over x_4^2}) \cr
&& + {8\hbar\over v_2^3\,x_1^2\,x_2^2\,x_3^2}\,({1\over x_1 x_2}+{1\over x_1 x_3}+{1\over x_1 x_4}+{1\over x_2 x_3}+{1\over x_2 x_4}+{1\over x_3 x_4}) \cr
&& -  {12\hbar v_3 \over v_2^4\,x_1^2\,x_2^2\,x_3^2}\,({1\over x_1}+{1\over x_2}+{1\over x_3}+{1\over x_4}) 
+ {12\hbar\, v_3^2\over v_2^5\,x_1^2\,x_2^2\,x_3^2}
- {6\hbar\, v_4\over v_2^4\,x_1^2\,x_2^2\,x_3^2}
\eea

\bea
W_1^{(1)} 
&=& {1\over \hbar x} + {1\over v_2\, x^3} - {v_3\over v_2^2\, x^2}
\eea

\bea
W_2^{(1)} 
&=& {3\over v_2^2\, x_1^2 x_2^2}\,({1\over x_1^2}+{1\over x_2^2}+{2\over 3\, x_1 x_2})
+ {1\over \hbar\,v_2\, x_1^2\,x_2^2} - {4 v_3\over v_2^3\, x_1^2 x_2^2}\,({1\over x_1}+{1\over x_2}) \cr
&& +{4 v_3^2\over v_2^4\, x_1^2 x_2^2}
-{3 v_4\over v_2^3\, x_1^2 x_2^2}
\eea

\bea
W_3^{(1)} 
&=& {12\over v_2^3\, x_1^2 x_2^2 x_3^2}\,({1\over x_1^3}+{1\over x_2^3}+{1\over x_3^3}) \cr
&& + {12\over v_2^3\, x_1^2 x_2^2 x_3^2}\,({1\over x_1^2 x_2}+{1\over x_2^2 x_3}+{1\over x_3^2 x_1}+{1\over x_1 x_2^2}+{1\over x_2 x_3^2}+{1\over x_3 x_1^2})\cr
&& + {8\over v_2^3\, x_1^3 x_2^3 x_3^3}  
+ {2\over \hbar v_2^2  x_1^2 x_2^2 x_3^2}\,({1\over x_1}+{1\over x_2}+{1\over x_3}) \cr
&& - {24 v_3\over v_2^4  x_1^2 x_2^2 x_3^2}\,({1\over x_1^2}+{1\over x_2^2}+{1\over x_3^2}+{1\over x_1 x_2}+{1\over x_2 x_3}+{1\over x_3 x_1})  
- {2 v_3\over \hbar v_2^3  x_1^2 x_2^2 x_3^2} \cr
&& + {32 v_3^2\over v_2^5  x_1^2 x_2^2 x_3^2} \,({1\over x_1}+{1\over x_2}+{1\over x_3})
- {32 v_3^3\over v_2^6  x_1^2 x_2^2 x_3^2}
- {18 v_4\over v_2^4  x_1^2 x_2^2 x_3^2}\,({1\over x_1}+{1\over x_2}+{1\over x_3})\cr
&& + {42 v_3 v_4\over v_2^5  x_1^2 x_2^2 x_3^2}
- {12 v_5\over v_2^4  x_1^2 x_2^2 x_3^2}
\eea

\bea
W_1^{(2)} 
&=& - {1\over \hbar^3 x} 
+ {3\over \hbar\, v_2^2\, x^5} 
- {5 v_3\over \hbar\, v_2^3\, x^4}
+ {5 v_3^2\over \hbar\, v_2^4\, x^3}
- {5 v_3^3\over \hbar\, v_2^5\, x^2}
- {3 v_4\over \hbar\, v_2^3\, x^3} \cr
&& + {8 v_3\, v_4\over \hbar\, v_2^4\, x^2}
- {3 v_5\over \hbar\, v_2^3\, x^2} \cr
\eea

\bea
W_2^{(2)} 
&=& {15\over \hbar\, v_2^3\, x_1^2 x_2^2}\,({1\over x_1^4}+{1\over x_2^4}+{1\over  x_1^2\, x_2^2})
+ {12\over \hbar\, v_2^3\, x_1^2 x_2^2}\,({1\over x_1^3 x_2}+{1\over x_1 x_2^3 })
- {1\over \hbar^3\, v_2\, x_1^2 x_2^2} \cr
&& - {32 v_3\over \hbar\, v_2^4\, x_1^2 x_2^2}\,({1\over x_1^3}+{1\over x_2^3})
 - {30 v_3\over \hbar\, v_2^4\, x_1^2 x_2^2}\,({1\over x_1 x_2^2}+{1\over x_1^2 x_2})
+ {45 v_3^2\over \hbar\, v_2^5\, x_1^2 x_2^2}\,({1\over x_1^2}+{1\over x_2^2})\cr
&& + {40 v_3^2\over \hbar\, v_2^5\, x_1^3 x_2^3} 
 - {50 v_3^3\over \hbar\, v_2^6\, x_1^2 x_2^2}\,({1\over x_1}+{1\over x_2})
+ {50 v_3^4\over \hbar\, v_2^7\, x_1^2 x_2^2}
- {24 v_4\over \hbar\, v_2^4\, x_1^2 x_2^2}\,({1\over x_1^2}+{1\over x_2^2}) \cr
&& - {18 v_4\over \hbar\, v_2^4\, x_1^3 x_2^3}
+ {64 v_3\, v_4\over \hbar\, v_2^5\, x_1^2 x_2^2}\,({1\over x_1}+{1\over x_2})
- {109 v_3^2\, v_4\over \hbar\, v_2^6\, x_1^2 x_2^2}
+ {24 v_4^2\over \hbar\, v_2^5\, x_1^2 x_2^2} \cr
&& - {18 v_5\over \hbar\, v_2^4\, x_1^2 x_2^2}\,({1\over x_1}+{1\over x_2}) 
 + {50 v_3 \, v_5\over \hbar\, v_2^5\, x_1^2 x_2^2}
- {15 v_6\over \hbar\, v_2^4\, x_1^2 x_2^2}
\eea

\bea
W_1^{(3)} 
&=&  {2\over \hbar^5 x} 
+ {15 \over \hbar^2\, v_2^3\, x^7}
- {3 \over \hbar^3\, v_2^2\, x^5} 
- {35 v_3\over \hbar^2\, v_2^4\, x^6}
+ {5 v_3\over \hbar^3\, v_2^3\, x^4}
+ {50 v_3^2\over \hbar^2\, v_2^5\, x^5}
- {5 v_3^2\over \hbar^3\, v_2^4\, x^3} \cr
&& - {60 v_3^3\over \hbar^2\, v_2^6\, x^4}
+ {5 v_3\over \hbar^3\, v_2^5\, x^2} 
+ {60 v_3^4\over \hbar^2\, v_2^7\, x^3}
- {60 v_3^5\over \hbar^2\, v_2^8\, x^2}
- {24 v_4\over \hbar^2\, v_2^4\, x^5}
+ {3 v_4\over \hbar^3\, v_2^3\, x^3} \cr
&& + {75 v_3 v_4\over \hbar^2\, v_2^5\, x^4} 
 - {8 v_3 v_4\over \hbar^3\, v_2^4\, x^2}
- {125 v_3^2 v_4\over \hbar^2\, v_2^6\, x^3}
+ {185 v_3^3 v_4\over \hbar^2\, v_2^7\, x^2}
+ {24 v_4^2\over \hbar^2\, v_2^5\, x^3}
- {99 v_3 v_4^2\over \hbar^2\, v_2^6\, x^2} \cr
&& - {21 v_5\over \hbar^2\, v_2^4\, x^4}
+ {3 v_5\over \hbar^3\, v_2^3\, x^2} 
 + {56 v_3 v_5\over \hbar^2\, v_2^5\, x^3}
- {106 v_3^2 v_5\over \hbar^2\, v_2^6\, x^2}
+ {45 v_4 v_5\over \hbar^2\, v_2^5\, x^2}
- {15 v_6\over \hbar^2\, v_2^4\, x^3} \cr
&& + {50 v_3 v_6\over \hbar^2\, v_2^5\, x^2}
- {15 v_7\over \hbar^2\, v_2^4\, x^2}
\eea

The free energies are:
\beq
F_1 = {1\over 2} \ln{(v_2/\hbar)}
\eeq

\beq
F_2 = -{5 v_3^2\over 6\hbar\, v_2^3} + {3 v_4\over 4\hbar\, v_2^2}
\eeq

\beq
F_3
= {5 v_3^2\over 6\hbar^3\,v_2^3}
- {5 v_3^4\over \hbar^2\,v_2^6}
- {3 v_4\over 4 \hbar^3\,v_2^2}
+ {25 v_3^2 v_4\over 2 \hbar^2\,v_2^5}
- {3 v_4^2\over \hbar^2\,v_2^4}
- {7 v_3 v_5\over \hbar^2\,v_2^4}
+ {5 v_6\over 2 \hbar^2\,v_2^3}
\eeq

\vfill\eject

\end{document}